\documentclass{cjps}
\usepackage{fancyhdr}
\usepackage{graphicx}
\usepackage{latexsym}
\usepackage{amssymb}
\usepackage{amsmath}
\usepackage{amsthm}
\usepackage{array}
\newcommand{\lsim}{\mathrel{\rlap{\lower4pt\hbox{\hskip0pt$\sim$}}
\raise1pt\hbox{$<$}}}           
\newcommand{\gsim}{\mathrel{\rlap{\lower4pt\hbox{\hskip0pt$\sim$}}
\raise1pt\hbox{$>$}}}           
\fancypagestyle{cjp}{ \fancyhf{} \fancyhead[L]{\small\textrm{CHINESE
JOURNAL OF PHYSICS}} \fancyhead[C]{\small\textrm{VOL. 49, NO. 5}}
\fancyhead[R]{\small\textrm{October 2011}}
\fancyfoot[L]{\small\textrm{http://PSROC.phys.ntu.edu.tw/cjp}}
\fancyfoot[C]{\small\textbf\thepage} \fancyfoot[R]{\small\textrm{
              $\copyright$ 2011 THE PHYSICAL SOCIETY \\
              OF THE REPUBLIC OF CHINA}}

}

\begin{document}
\setcounter{page}{955}          

\vspace*{15pt}
\title{Selected Highlights from the Study of Mesons}

\author{Lei Chang}
\affiliation{Physics Division, Argonne National Laboratory, Argonne, Illinois 60439, USA}

\author{Craig D.~Roberts}
\affiliation{Physics Division, Argonne National Laboratory, Argonne, Illinois 60439, USA}
\affiliation{Institut f\"ur Kernphysik, Forschungszentrum J\"ulich, D-52425 J\"ulich, Germany}
\affiliation{Department of Physics, Center for High Energy Physics and State Key Laboratory of Nuclear Physics and Technology, Peking University, Beijing 100871, China}
\affiliation{Department of Physics, Illinois Institute of Technology, Chicago, Illinois 60616-3793, USA}

\author{Peter~C.~Tandy}
\affiliation{Center for Nuclear Research, Department of Physics, Kent State University, Kent OH 44242, USA}

\received{July 20, 2011}

\begin{abstract}
We provide a brief review of recent progress in the study of mesons using QCD's Dyson-Schwinger equations.  Along the way we touch on aspects of confinement and dynamical chiral symmetry breaking but in the main focus upon:
exact results for pseudoscalar mesons, including aspects of the $\eta$-$\eta^\prime$ problem;
a realisation that the so-called vacuum condensates are actually an intrinsic, localised property of hadrons;
an essentially nonperturbative procedure for constructing a symmetry-preserving Bethe-Salpeter kernel, which has enabled a demonstration that dressed-quarks possess momentum-dependent anomalous chromo- and electromagnetic moments that are large at infrared momenta, and resolution of a longstanding problem in understanding the mass-splitting between $\rho$- and $a_1$-mesons such that they are now readily seen to be parity partners in the meson spectrum;
features of electromagnetic form factors connected with charged and neutral pions;
and computation and explanation of valence-quark distribution functions in pseudoscalar mesons.
We argue that in solving QCD, a constructive feedback between theory and extant and forthcoming experiments will enable constraints to be placed on the infrared behaviour of QCD's $\beta$-function, the nonperturbative quantity at the core of hadron physics.
\end{abstract}



\maketitle
\thispagestyle{cjp}



\section{Introduction}
A hundred years and more of fundamental research in atomic and nuclear physics has shown that all matter is corpuscular, with the atoms that comprise us, themselves containing a dense nuclear core.  This core is composed of protons and neutrons, referred to collectively as nucleons, which are members of a broader class of femtometre-scale particles, called hadrons.  In working toward an understanding of hadrons, we have discovered that they are complicated bound-states of quarks and gluons.  These quarks and gluons are elementary, pointlike excitations, whose interactions are described by a Poincar\'e invariant quantum non-Abelian gauge field theory; namely, quantum chromodynamics (QCD).  The goal of hadron physics is the provision of a quantitative explanation of the properties of hadrons through a solution of QCD.

Quantum chromodynamics is the strong-interaction part of the Standard Model of Particle Physics and solving QCD presents a fundamental problem that is unique in the history of science.  Never before have we been confronted by a theory whose elementary excitations are not those degrees-of-freedom readily accessible via experiment; i.e., whose elementary excitations are \emph{confined}.  Moreover, there are numerous reasons to believe that QCD generates forces which are so strong that less-than 2\% of a nucleon's mass can be attributed to the so-called current-quark masses that appear in QCD's Lagrangian; viz., forces capable of generating mass from nothing, a phenomenon known as dynamical chiral symmetry breaking (DCSB).

Neither confinement nor DCSB is apparent in QCD's Lagrangian and yet they play the dominant role in determining the observable characteristics of real-world QCD.  The physics of hadrons is ruled by \emph{emergent phenomena} such as these, which can only be elucidated through the use of nonperturbative methods in quantum field theory.  This is both the greatest novelty and the greatest challenge within the Standard Model.  We must find essentially new ways and means to explain precisely via mathematics the observable content of QCD.

The complex of Dyson-Schwinger equations (DSEs) is a powerful tool, which has been employed with marked success to study confinement and DCSB, and their impact on hadron observables.  This will be emphasised and exemplified in this concise review, which describes selected recent progress in the study of mesons.  It complements and extends earlier and other efforts \cite{Roberts:1994dr,Roberts:2000aa,Maris:2003vk,Pennington:2005be,Holl:2006ni,%
Fischer:2006ub,Roberts:2007jh,Roberts:2007ji,Holt:2010vj,Swanson:2010pw}.

\section{Hadron Physics}
\label{sect:HP}
The basic problem of hadron physics is to solve QCD.  This inspiring goal will only be achieved through a joint effort from experiment and theory because it is the feedback between them that leads most rapidly to improvements in understanding.  The hadron physics community now has a range of major facilities that are accumulating data, of unprecedented accuracy and precision, which pose important challenges for theory.  The opportunities for researchers in hadron physics promise to expand with the use of extant accelerators, and upgraded and new machines and detectors that will appear on a five-to-ten-year time-scale, in China, Germany, Japan, Switzerland and the USA.
A short list of facilities may readily be compiled: Beijing's electron-positron collider; in Germany --
COSY (J\"ulich Cooler Synchrotron),
ELSA (Bonn Electron Stretcher and Accelerator),
MAMI (Mainz Microtron), and
FAIR (Facility for Antiproton and Ion Research) under construction near Darmstadt;
in Japan -- J-PARC (Japan Proton Accelerator Research Complex) under construction in Tokai-Mura, 150km NE of Tokyo, and
KEK, Tsukuba;
in Switzerland, the ALICE and COMPASS detectors at CERN;
and in the USA, both the Thomas Jefferson National Accelerator Facility (JLab), currently being upgraded, with new generation experiments expected in 2016, and RHIC (Relativistic Heavy Ion Collider) at Brookhaven National Laboratory.  

Asymptotic coloured states have not been observed, but is it a cardinal fact that they cannot?  No solution to QCD will be complete if it does not explain confinement.  This means confinement in the real world, which contains quarks with light current-quark masses.  This is distinct from the artificial universe of pure-gauge QCD without dynamical quarks, studies of which tend merely to focus on achieving an area law for a Wilson loop and hence are irrelevant to the question of light-quark confinement.

In stepping toward an answer to the question of confinement, it will likely be necessary to map out the long-range behaviour of the interaction between light-quarks; namely, QCD's $\beta$-function at infrared momenta.  In this connection it is noteworthy that the spectrum of meson and baryon excited states, and hadron elastic and transition form factors provide unique information about the long-range interaction between light-quarks and, in addition, the distribution of a hadron's characterising properties -- such as mass and momentum, linear and angular -- amongst its QCD constituents.  The upgraded and promised future facilities will provide data that should guide the charting process.   However, to make full use of that data, it will be necessary to have Poincar\'e covariant theoretical tools that enable the reliable study of hadrons in the mass range $1$-$2\,$GeV.  Crucially, on this domain both confinement and DCSB are germane.

\begin{figure}[t]

\centerline{
\includegraphics[clip,width=0.66\textwidth]{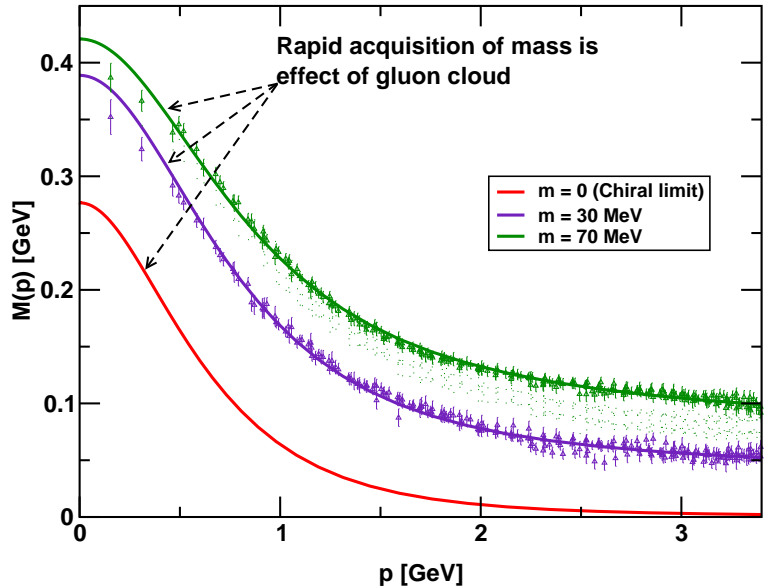}}

\caption{\label{gluoncloud}
Dressed-quark mass function, $M(p)$ in Eq.\,(\protect\ref{SgeneralN}): \emph{solid curves} -- DSE results, explained in \protect\cite{Bhagwat:2003vw,Bhagwat:2006tu}, ``data'' -- numerical simulations of lattice-QCD \protect\cite{Bowman:2005vx}.  (NB.\ $m=70\,$MeV is the uppermost curve and current-quark mass decreases from top to bottom.)  One observes the current-quark of perturbative QCD evolving into a constituent-quark as its momentum becomes smaller.  The constituent-quark mass arises from a cloud of low-momentum gluons attaching themselves to the current-quark.  This is dynamical chiral symmetry breaking (DCSB): an essentially nonperturbative effect that generates a quark mass \emph{from nothing}; namely, it occurs even in the chiral limit.  
(Figure adapted from Ref.\,\protect\cite{Bhagwat:2007vx}.)}
\end{figure}

It is known that DCSB; namely, the generation of mass \emph{from nothing}, does take place in QCD.  It arises primarily because a dense cloud of gluons comes to clothe a low-momentum quark \cite{Bhagwat:2007vx}.  This is readily seen by solving the DSE for the dressed-quark propagator; i.e., the gap equation, which yields the result illustrated in Fig.\,\ref{gluoncloud}.  (In our Euclidean metric:  $\{\gamma_\mu,\gamma_\nu\} = 2\delta_{\mu\nu}$; $\gamma_\mu^\dagger = \gamma_\mu$; $\gamma_5= \gamma_4\gamma_1\gamma_2\gamma_3$, tr$[\gamma_4\gamma_\mu\gamma_\nu\gamma_\rho\gamma_\sigma]=-4 \epsilon_{\mu\nu\rho\sigma}$; $\sigma_{\mu\nu}=(i/2)[\gamma_\mu,\gamma_\nu]$; $a \cdot b = \sum_{i=1}^4 a_i b_i$; and $P_\mu$ timelike $\Rightarrow$ $P^2<0$.  More information is available in App.\,A of Ref.\,\protect\cite{Holl:2006ni}.)
However, the origin of the interaction strength at infrared momenta, which guarantees DCSB through the gap equation, is currently unknown.  This relationship ties confinement to DCSB.  The reality of DCSB means the Higgs mechanism is largely irrelevant to the bulk of normal matter in the Universe.  Instead the most important mass generating mechanism for light-quark hadrons is the strong interaction effect of DCSB; e.g., one may identify it as being responsible for 98\% of a proton's mass \cite{Flambaum:2005kc,Holl:2005st}.
There is a caveat; namely, as so often, the pion is exceptional.  Its mass is given by the simple product of two terms, one of which is the ratio of two order parameters for DCSB whilst the other is determined by the current-quark mass (Sec.\,\ref{psmassformula}).  Hence the pion would be massless in the absence of a mechanism that can generate a current-mass for at least one light-quark.  The impact of a massless, strongly-interacting particle on the physics of the Universe would be dramatic.

It is natural to ask whether the connection between confinement and DCSB is accidental or causal.  There are models with DCSB but not confinement, however, a model with confinement but lacking DCSB has not yet been identified (see, e.g., Secs.\,2.1 and 2.2 of Ref.\,\cite{Roberts:2007jh}).  This leads to a conjecture that DCSB is a necessary consequence of confinement.  It is interesting that there are numerous models and theories which exhibit both confinement and DCSB, and possess an external control parameter such that deconfinement and chiral symmetry restoration occur simultaneously at some critical value of this parameter; e.g., quantum electrodynamics in three dimensions with $N_f$ electrons \cite{Bashir:2008fk,Bashir:2009fv,Hofmann:2010zy}, and models of QCD at nonzero temperature and chemical potential \cite{Bender:1996bm,Blaschke:1997bj,Bender:1997jf,Chen:2008zr,Fischer:2009gk,Qin:2010nq,%
Qin:2010pc,Liu:2011zz,Qin:2011zz}.  Whether this simultaneity is a property possessed by QCD, and/or some broader class of theories, in response to changes in: the number of light-quark flavours; temperature; or chemical potential, is a longstanding question.

The momentum-dependence of the quark mass, illustrated in Fig.\,\ref{gluoncloud}, is an essentially quantum field theoretic effect, unrealisable in quantum mechanics, and a fundamental feature of QCD.  This single curve connects the infrared and ultraviolet regimes of the theory, and establishes that the constituent-quark and current-quark masses are simply two connected points separated by a large momentum interval.  The curve shows that QCD's dressed-quark behaves as a constituent-quark, a current-quark, or something in between, depending on the momentum of the probe which explores the bound-state containing the dressed-quark.  It follows that calculations addressing momentum transfers $Q^2 \gsim M^2$, where $M$ is the mass of the hadron involved, require a Poincar\'e-covariant approach that can veraciously realise quantum field theoretical effects \cite{Cloet:2008re}.  Owing to the vector-exchange character of QCD, covariance also guarantees the existence of nonzero quark orbital angular momentum in a hadron's rest-frame \cite{Bhagwat:2006xi,Bhagwat:2006pu,Cloet:2007pi,Roberts:2007ji}.

The dressed-quark mass function has a remarkable capacity to correlate and to contribute significantly in explaining a wide range of diverse phenomena.  This brings urgency to the need to understand the relationship between parton properties in the light-front frame, whose peculiar properties simplify some theoretical analyses, and the structure of hadrons as measured in the rest frame or other smoothly related frames.  This is a problem because, e.g., DCSB, an established keystone of low-energy QCD, has not explicitly been realised in the light-front formulation.  The obstacle is the constraint $k^+:=k^0+k^3>0$ for massive quanta on the light front \cite{Brodsky:1991ir}.  It is therefore impossible to make zero momentum Fock states that contain particles and hence the vacuum is ``trivial''.
On the other hand, it is conceivable that DCSB is inextricably tied with the formation and structure of Goldstone modes and not otherwise a measurable property of the vacuum.  This conjecture is being explored \cite{Brodsky:2010xf,Brodsky:2009zd,Chang:2011mu} and is something about which more will be written herein (Sec.\,\ref{inhadroncondensate}).
In addition, parton distribution functions, which have a probability interpretation in the infinite momentum frame, must be calculated in order to comprehend their content: parametrisation is insufficient.  It would be very interesting to know, e.g., how, if at all, the distribution functions of a Goldstone mode differ from those of other hadrons \cite{Holt:2010vj}.

\section{Confinement}
\label{Sect:Conf}
It is worth stating plainly that the potential between infinitely-heavy quarks measured in numerical simulations of quenched lattice-regularised QCD -- the so-called static potential -- is simply \emph{irrelevant} to the question of confinement in the real world, in which light quarks are ubiquitous.  In fact, it is a basic feature of QCD that light-particle creation and annihilation effects are essentially nonperturbative and therefore it is impossible in principle to compute a potential between two light quarks \cite{Bali:2005fu,Chang:2009ae}.

Drawing on a long list of sources; e.g., Refs.\,\cite{Gribov:1999ui,Munczek:1983dx,Stingl:1983pt,Cahill:1988zi}, a perspective on confinement was laid out in Ref.\,\cite{Krein:1990sf}.  Confinement can be related to the analytic properties of QCD's Schwinger functions, which are often called Euclidean-space Green functions.  For example, it can be read from the reconstruction theorem \cite{SW80,GJ81} that the only Schwinger functions which can be associated with expectation values in the Hilbert space of observables; namely, the set of measurable expectation values, are those that satisfy the axiom of reflection positivity.  This is an extremely tight constraint.  It can be shown to require as a necessary condition that the Fourier transform of the momentum-space Schwinger function is a positive-definite function of its arguments.  This condition suggests a practical confinement test, which can be used with numerical solutions of the DSEs (see, e.g., Sec.\,III.C of Ref.\,\cite{Hawes:1993ef} and Sec.\,IV of Ref.\,\cite{Maris:1995ns}).  The implications and use of reflection positivity are discussed and illustrated in Sec.~2 of Ref.\,\cite{Roberts:2007ji}.

It is noteworthy that any 2-point Schwinger function with an inflexion point at $p^2 > 0$ must breach the axiom of reflection positivity, so that a violation of positivity can be determined by inspection of the pointwise behaviour of the Schwinger function in momentum space (Sec.\,IV.B of Ref.\,\cite{Bashir:2008fk}).

\begin{figure}[t]

\centerline{
\includegraphics[clip,width=0.67\textwidth]{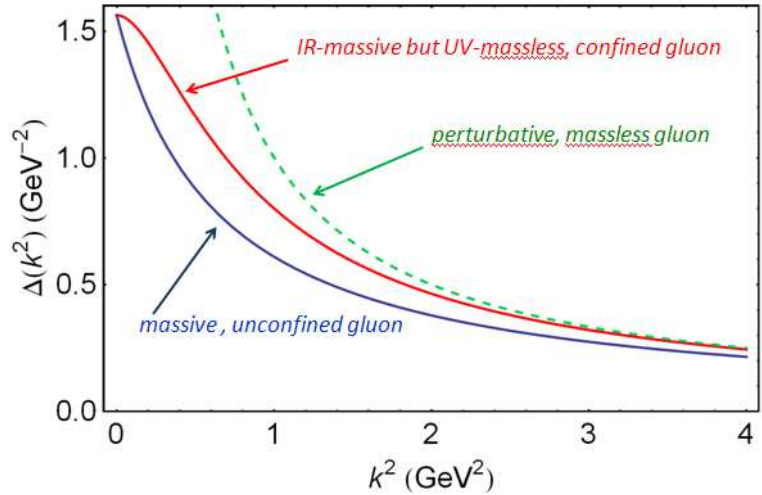}}

\caption{\label{fig:gluonrp}
$\Delta(k^2)$, the function that describes dressing of a Landau-gauge gluon propagator, plotted for three distinct cases.
A bare gluon is described by $\Delta(k^2) = 1/k^2$ (the dashed line), which is plainly convex on $k^2\in (0,\infty)$.  Such a propagator has a representation in terms of a non-negative spectral density.
In some theories, interactions generate a mass in the transverse part of the gauge-boson propagator, so that $\Delta(k^2) = 1/(k^2+m_g^2)$, which can also be represented in terms of a non-negative spectral density.
In QCD, however, self-interactions generate a momentum-dependent mass for the gluon, which is large at infrared momenta but vanishes in the ultraviolet \protect\cite{pepe:11}.  This is illustrated by the curve labelled ``IR-massive but UV-massless.''  With the generation of a mass-\emph{function}, $\Delta(k^2)$ exhibits an inflexion point and hence cannot be expressed in terms of a non-negative spectral density.
}
\end{figure}

Consider then $\Delta(k^2)$, which is the single scalar function that describes the dressing of a Landau-gauge gluon propagator.  A large body of work has focused on exposing the behaviour of $\Delta(k^2)$ in the pure Yang-Mills sector of QCD.  These studies are reviewed in Ref.\,\cite{pepe:11}.
A connection with the expression and nature of confinement in the Yang-Mills sector is indicated in Fig.\,\ref{fig:gluonrp}.  The appearance of an inflexion point in the two-point function generated by the gluon's momentum-dependent mass-function is impossible to overlook.  Hence this gluon cannot appear in the Hilbert space of observable states.  The inflexion point possessed by $M(p^2)$, visible in Fig.\,\ref{gluoncloud}, conveys the same properties on the dressed-quark propagator.

Numerical simulations of lattice-QCD confirm the appearance of an inflexion point in both the dressed-gluon and -quark propagators; e.g., see Fig.\,\ref{gluoncloud} and Ref.\,\cite{pepe:11}.  The signal is clearest for the gluon owing to the greater simplicity of simulations in the pure Yang-Mills sector \cite{Bonnet:2000kw,Skullerud:2000un,Kamleh:2007ud}.  We emphasise that this sense of confinement is essentially quantum field theoretical in nature.  Amongst its many consequences is that confinement in QCD cannot veraciously be represented in potential models.

{F}rom the perspective that confinement can be related to the analytic properties of QCD's Schwinger functions, the question of light-quark confinement can be translated into the challenge of charting the infrared behavior of QCD's \emph{universal} $\beta$-function.  (Although this function may depend on the scheme chosen to renormalise the theory, it is unique within a given scheme \protect\cite{Celmaster:1979km}.  Of course, the behaviour of the $\beta$-function on the perturbative domain is well known.)  This is a well-posed problem whose solution is an elemental goal of modern hadron physics and which can be addressed in any framework enabling the nonperturbative evaluation of renormalisation constants.  It is the $\beta$-function that is responsible for the behaviour evident in Figs.\,\ref{gluoncloud} and \ref{fig:gluonrp}, and one of the more interesting of contemporary questions is whether it is possible to reconstruct the $\beta$-function, or at least constrain it tightly, given empirical information on the gluon and quark mass functions.

\section{Gap and Bethe-Salpeter Equations}
\label{gapbse}
In order to proceed it is necessary to describe explicitly the best known and simplest DSE.  The Dyson or gap equation describes how quark propagation is influenced by interactions; viz., for a quark of flavour $f$,
\begin{equation}
 S_f(p)^{-1} = Z_2 \,(i\gamma\cdot p + m_f^{\rm bm}) + Z_1 \int^\Lambda_q\!\! g^2 D_{\mu\nu}(p-q)\frac{\lambda^a}{2}\gamma_\mu S_f(q) \frac{\lambda^a}{2}\Gamma^f_\nu(q,p) ,
\label{gendseN}
\end{equation}
where: $D_{\mu\nu}$ is the gluon propagator; $\Gamma^f_\nu$, the quark-gluon vertex; $\int^\Lambda_q$, a symbol that represents a Poincar\'e invariant regularization of the four-dimensional Euclidean integral, with $\Lambda$ the regularization mass-scale; $m_f^{\rm bm}(\Lambda)$, the current-quark bare mass; and $Z_{1,2}(\zeta^2,\Lambda^2)$, respectively, the vertex and quark wave function renormalisation constants, with $\zeta$ the renormalisation point -- dependence upon which is not usually made explicit.

The gap equation's solution is the dressed-quark propagator,
\begin{equation}
 S(p) =
%
\frac{1}{i \gamma\cdot p \, A(p^2,\zeta^2) + B(p^2,\zeta^2)}
= \frac{Z(p^2,\zeta^2)}{i\gamma\cdot p + M(p^2)}\,,
%
\label{SgeneralN}
\end{equation}
which is obtained from Eq.\,(\ref{gendseN}) augmented by a renormalisation condition.  A mass-independent scheme is a useful choice and can be implemented by fixing all renormalisation constants in the chiral limit.  (See, e.g., Ref.\,\cite{Chang:2008ec} and references therein; or Ref.\,\protect\cite{tarrach} for a detailed discussion of renormalisation.)

The mass function, $M(p^2)=B(p^2,\zeta^2)/A(p^2,\zeta^2)$, is independent of the renormalisation point, $\zeta$; and the renormalised current-quark mass,
\begin{equation}
\label{mzeta}
m_f^\zeta = Z_m(\zeta,\Lambda) \, m^{\rm bm}(\Lambda) = Z_4^{-1} Z_2\, m_f^{\rm bm},
\end{equation}
wherein $Z_4$ is the renormalisation constant associated with the Lagrangian's mass-term. Like the running coupling constant, this ``running mass'' is familiar from textbooks.  However, it is not commonly appreciated that $m^\zeta$ is simply the dressed-quark mass function evaluated at one particular deep spacelike point; viz,
\begin{equation}
m_f^\zeta = M_f(\zeta^2)\,.
\end{equation}
The renormalisation-group invariant current-quark mass may be inferred via
\begin{equation}
\hat m_f = \lim_{p^2\to\infty} \left[\frac{1}{2}\ln \frac{p^2}{\Lambda^2_{\rm QCD}}\right]^{\gamma_m} M_f(p^2)\,,
\end{equation}
where $\gamma_m = 12/(33-2 N_f)$: $N_f$ is the number of quark flavours employed in computing the running coupling; and $\Lambda_{\rm QCD}$ is QCD's dynamically-generated renormalisation-group-invariant mass-scale.  The chiral limit is expressed by
\begin{equation}
\hat m_f = 0\,.
\end{equation}
Moreover,
\begin{equation}
\forall \zeta \gg \Lambda_{\rm QCD}, \;
\frac{m_{f_1}^\zeta}{m^\zeta_{f_2}}=\frac{\hat m_{f_1}}{\hat m_{f_2}}\,.
\end{equation}
However, we would like to emphasise that in the presence of DCSB the ratio $m_{f_1}^{\zeta=p^2}/m^{\zeta=p^2}_{f_2}=M_{f_1}(p^2)/M_{f_2}(p^2)$ is not independent of $p^2$: in the infrared; i.e., $\forall p^2 \lesssim \Lambda_{\rm QCD}^2$, it then expresses a ratio of constituent-like quark masses, which, for light quarks, are two orders-of-magnitude larger than their current-masses and nonlinearly related to them \cite{Flambaum:2005kc,Holl:2005st}.

The gap equation illustrates the features and flaws of each DSE.  It is a nonlinear
integral equation for the dressed-quark propagator and hence can yield much-needed nonperturbative information.  However, the kernel involves the two-point function $D_{\mu\nu}$ and the three-point function $\Gamma^f_\nu$.  The gap equation is therefore coupled to the DSEs satisfied by these functions, which in turn involve higher $n$-point functions.  Hence the DSEs are a tower of coupled integral equations, with a tractable problem obtained only once a truncation scheme is specified.  It is unsurprising that the best known truncation scheme is the weak coupling expansion, which reproduces every
diagram in perturbation theory.  This scheme is systematic and valuable in the analysis
of large momentum transfer phenomena because QCD is asymptotically free but it precludes any possibility of obtaining nonperturbative information.

Given the importance of DCSB in QCD, it is significant that the dressed-quark propagator features in the axial-vector Ward-Takahashi identity, which expresses chiral symmetry and its breaking pattern:
\begin{equation}
P_\mu \Gamma_{5\mu}^{fg}(k;P) + \, i\,[m_f(\zeta)+m_g(\zeta)] \,\Gamma_5^{fg}(k;P)
= S_f^{-1}(k_+) i \gamma_5 +  i \gamma_5 S_g^{-1}(k_-) \,,
\label{avwtimN}
\end{equation}
where $P=p_1+p_2$ is the total-momentum entering the vertex and $k$ is the relative-momentum between the amputated quark legs.  To be explicit, $k=(1-\eta) p_1 + \eta p_2$, with $\eta \in [0,1]$, and hence $k_+ = p_1 = k + \eta P$, $k_- = p_2 = k - (1-\eta) P$.  In a Poincar\'e covariant approach, such as presented by a proper use of DSEs, no observable can depend on $\eta$; i.e., the definition of the relative momentum.  NB.\ Sec.\,\ref{flavourless} discusses the important differences encountered in treating flavourless pseudoscalar mesons.

In Eq.\,(\ref{avwtimN}), $\Gamma_{5\mu}^{fg}$ and $\Gamma_5^{fg}$ are, respectively, the amputated axial-vector and pseudoscalar vertices.  They are both obtained from an inhomogeneous Bethe-Salpeter equation (BSE), which is exemplified here using a textbook expression \cite{Salpeter:1951sz}:
\begin{equation}
[\Gamma_{5\mu}(k;P)]_{tu} = Z_2 [\gamma_5 \gamma_\mu]_{tu}+ \int_q^\Lambda [ S(q_+) \Gamma_{5\mu}(q;P) S(q_-) ]_{sr} K_{tu}^{rs}(q,k;P),
\label{bsetextbook}
\end{equation}
in which $K$ is the fully-amputated quark-antiquark scattering kernel, and
the colour-, Dirac- and flavour-matrix structure of the elements in the equation is  denoted by the indices $r,s,t,u$.  NB.\ By definition, $K$ does not contain quark-antiquark to single gauge-boson annihilation diagrams, nor diagrams that become disconnected by cutting one quark and one antiquark line.

The Ward-Takahashi identity, Eq.\,(\ref{avwtimN}), entails that an intimate relation exists between the kernel in the gap equation and that in the BSE.  (This is another example of the coupling between DSEs.)  Therefore an understanding of chiral symmetry and its dynamical breaking can only be obtained with a truncation scheme that preserves this relation, and hence guarantees Eq.\,(\ref{avwtimN}) without a fine-tuning of model-dependent parameters.

\section{Nonperturbative Truncation}
\label{spectrum1}
Through the gap and Bethe-Salpeter equations the pointwise behaviour of the $\beta$-function determines the pattern of chiral symmetry breaking; e.g., the behaviour in Fig.\,\ref{gluoncloud}.  Moreover, the fact that these and other DSEs connect the $\beta$-function to experimental observables entails that comparison between computations and observations of the hadron mass spectrum, and hadron elastic and transition form factors, can be used to constrain the $\beta$-function's long-range behaviour.

In order to realise this goal, a nonperturbative symmetry-preserving DSE truncation is necessary.  Steady quantitative progress can be made with a scheme that is systematically improvable \cite{Munczek:1994zz,Bender:1996bb}.  In fact, the mere existence of such a scheme has enabled the proof of exact nonperturbative results in QCD.

Before describing a number of these in some detail, it is worth explicating the range of applications.  For example, there are:
veracious statements about the pion $\sigma$-term \cite{Flambaum:2005kc};
radially-excited and hybrid pseudoscalar mesons \cite{Holl:2004fr,Holl:2005vu};
heavy-light \cite{Ivanov:1998ms} and heavy-heavy mesons \cite{Bhagwat:2006xi};
novel results for the pion susceptibility obtained via analysis of the isovector-pseudoscalar vacuum polarisation \cite{Chang:2009at}, which bear upon the essential content of the so-called ``Mexican hat'' potential that is used in building models for QCD;
and a derivation \cite{Chang:2008sp} of the Weinberg sum rule \cite{Weinberg:1967kj}.

\subsection{Pseudoscalar meson mass formula}
\label{psmassformula}
Turning now to a fuller illustration, the first of the results was introduced in Ref.\,\cite{Maris:1997hd}; namely, a mass formula that is exact for flavour non-diagonal pseudoscalar mesons:
\begin{equation}
\label{mrtrelation}
f_{H_{0^-}} m_{H_{0^-}}^2 = (m_{f_1}^\zeta + m_{f_2}^\zeta) \rho_{H_{0^-}}^\zeta,
\end{equation}
where: $m_{f_i}^\zeta$ are the current-masses of the quarks constituting the meson; and
\begin{eqnarray}
f_{H_{0^-}} P_\mu = \langle 0 | \bar q_{f_2} \gamma_5 \gamma_\mu q_{f_1} |H_{0^-}\rangle
& = & Z_2\; {\rm tr}_{\rm CD}
\int_q^\Lambda i\gamma_5\gamma_\mu S_{f_1}(q_+) \Gamma_{H_{0^-}}(q;P) S_{f_2}(q_-)\,, \label{fpigen}
\\
i\rho_{H_{0^-}} = -\langle 0 | \bar q_{f_2} i\gamma_5 q_{f_1} |H_{0^-} \rangle & = & Z_4\; {\rm tr}_{\rm CD}
\int_q^\Lambda \gamma_5 S_{f_1}(q_+) \Gamma_{H_{0^-}}(q;P) S_{f_2}(q_-) \,,
\label{rhogen}
\end{eqnarray}
where $\Gamma_{H_{0^-}}$ is the pseudoscalar meson's bound-state Bethe-Salpeter amplitude:
\begin{eqnarray}
\nonumber \Gamma_{H_{0^-}}(k;P) &=& \gamma_5 \left[ i E_{H_{0^-}}(k;P) + \gamma\cdot P F_{H_{0^-}}(k;P) \right. \\
&& \left. + \gamma\cdot k \, G_{H_{0^-}}(k;P) - \sigma_{\mu\nu} k_\mu P_\nu H_{H_{0^-}}(k;P) \right],
\label{genGpi}
\end{eqnarray}
which is determined from the associated homogeneous BSE.

It is worth emphasising that the quark wavefunction and Lagrangian mass renormalisation constants, $Z_{2,4}(\zeta,\Lambda)$, respectively, depend on the gauge parameter in precisely the manner needed to ensure that the right-hand sides of Eqs.\,(\ref{fpigen}), (\ref{rhogen}) are gauge-invariant.  Moreover, $Z_2(\zeta,\Lambda)$ ensures that the right-hand side of Eq.\,(\ref{fpigen}) is independent of both $\zeta$ and $\Lambda$, so that $f_{H_{0^-}}$ is truly an observable; and $Z_4(\zeta,\Lambda)$ ensures that $\rho_{H_{0^-}}^\zeta$ is independent of $\Lambda$ and evolves with $\zeta$ in just the way necessary to guarantee that the product $m^\zeta \rho_{H_{0^-}}^\zeta$ is
renormalisation-point-independent.  In addition, it should be noted that Eq.\,(\ref{mrtrelation}) is valid for every pseudoscalar meson and for any value of the current-quark masses; viz., $\hat m_{f_i} \in [ 0,\infty)$, $i=1,2$.  This includes arbitrarily large values and also the chiral limit, in whose neighbourhood Eq.\,(\ref{mrtrelation}) can be shown \cite{Maris:1997hd} to reproduce the familiar Gell-Mann--Oakes--Renner relation.

The axial-vector Ward-Takahashi identity, Eq.\,(\ref{avwtimN}), is a crucial bridge to Eqs.\,(\ref{mrtrelation}) -- (\ref{rhogen}); and on the way one can also prove the following Goldberger-Treiman-like relations \cite{Maris:1997hd}:
\begin{eqnarray}
\label{gtlrelE}
f_{H_{0^-}}^0 E_{H_{0^-}}(k;0) &=& B^0(k^2)\,,\\
\label{gtlrelF}
F_R(k;0) + 2 f_{H_{0^-}}^0 F_{H_{0^-}}(k;0) &=& A^0(k^2)\,,\\
\label{gtlrelG}
G_R(k;0) + 2 f_{H_{0^-}}^0 G_{H_{0^-}}(k;0) &=& \frac{d}{dk^2}A^0(k^2)\,,\\
\label{gtlrelH}
H_R(k;0) + 2 f_{H_{0^-}}^0 H_{H_{0^-}}(k;0) &=& 0\,,
\end{eqnarray}
wherein the superscript indicates that the associated quantity is evaluated in the chiral limit, and $F_R$, $G_R$, $H_R$ are analogues in the inhomogeneous axial-vector vertex of the scalar functions in the $H_{0^-}$-meson's Bethe-Salpeter amplitude.

These identities are of critical importance in QCD.
The first, Eq.\,(\ref{gtlrelE}), can be used to prove that a massless pseudoscalar meson appears in the chiral-limit spectrum if, and only if, chiral symmetry is dynamically broken.  Moreover, it exposes the fascinating consequence that the solution of the two-body pseudoscalar bound-state problem is almost completely known once the one-body problem is solved for the dressed-quark propagator, with the relative momentum within the bound-state identified unambiguously with the momentum of the dressed-quark.  This latter emphasises that Goldstone's theorem has a pointwise expression in QCD.
The remaining three identities are also important because they show that a pseudoscalar meson \emph{must} contain components of pseudovector origin.  This result overturned a misapprehension of twenty-years standing; namely, that only $E_{H_{0^-}}(k;0)$ is nonzero \cite{delbourgoscadron}.  These pseudovector components materially influence the observable properties of pseudoscalar mesons \cite{Maris:1998hc,GutierrezGuerrero:2010md,Roberts:2010rn,Roberts:2011wy,Nguyen:2011jy}, as do their analogues in other mesons \cite{Maris:1999nt,Maris:1999bh,Maris:2000sk}.

It is natural to reiterate here a prediction for the properties of non-ground-state pseudoscalar mesons, which follows from the exact results described above; namely, in the chiral limit \cite{Holl:2004fr,Holl:2005vu}
\begin{equation}
\label{fpin}
f_{\pi_n} \equiv 0 \,, \forall n\geq 1\,.
\end{equation}
This is the statement that Goldstone modes are the only pseudoscalar mesons to possess a nonzero leptonic decay constant in the chiral limit when chiral symmetry is dynamically broken.  The decay constants of all other pseudoscalar mesons on this trajectory, e.g., radial excitations, vanish.  On the other hand, in the absence of DCSB the leptonic decay constant of each such pseudoscalar meson vanishes in the chiral limit; i.e, Eq.\,(\ref{fpin}) is true $\forall n \geq 0$.

\begin{figure}[t]

\centerline{
\includegraphics[clip,width=0.66\textwidth]{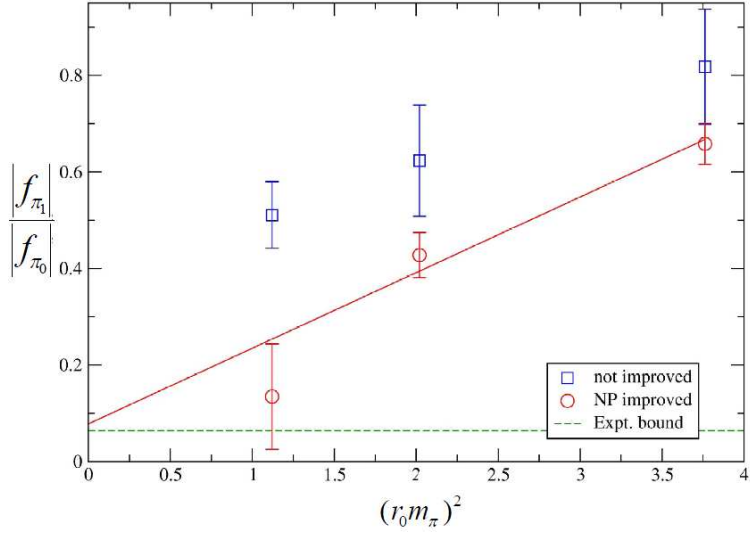}}

\caption{\label{fig:fRadial}
Lattice-QCD results for the ratio of the decay constants for the first-excited- and ground-state pseudoscalar mesons as a function of the pion mass squared.  (Lattice parameters: volume\,$=16^3\times 32$; $\beta = 5.2$, spacing $a\simeq 0.1\,$fm, two flavours of degenerate sea quarks; Wilson gauge action and clover fermions.)
The ``not improved'' results were obtained from a fermion action with poor chiral symmetry properties.  In this case $|f_{\pi_1}/f_{\pi_0}|\approx 0.4$, consistent with expectations based on quantum mechanics.
The ``improved'' results were obtained through implementation of the full ALPHA method for the nonperturbative improvement of the fermion action, which greatly improves the simulation's chiral symmetry properties.  In this case, $|f_{\pi_1}/f_{\pi_0}|\approx 0.01$.  (NB. The sign of the ratio was not determined in the simulation but is discussed in  Ref.\,\protect\cite{Qin:2011xq}.  Figure adapted from Ref.\,\protect\cite{McNeile:2006qy}.)}
\end{figure}

{F}rom the perspective of quantum mechanics, Eq.\,(\ref{fpin}) is a surprising fact.  The leptonic decay constant for $S$-wave states is typically proportional to the wave function at the origin.  Compared with the ground state, this is smaller for
an excited state because the wave function is broader in configuration space and wave functions are normalised.  However, it is a modest effect; e.g., consider the $e^+e^-$ decay of vector mesons, for which a calculation in relativistic quantum mechanics based on light-front dynamics \cite{deMelo:2005cy} yields $|f_{\rho_1}/f_{\rho_0}| = 0.5$, consistent with the value inferred from experiment.  Thus, it is not uncommon for Eq.\,(\ref{fpin}) to be perceived as ``remarkable'' or ``unbelievable.''  Notwithstanding this, in connection with the pion's first radial excitation, the value of $f_{\pi_1}= -2\,$MeV predicted in Ref.\,\cite{Holl:2004fr} is consistent with experiment \cite{Diehl:2001xe} and simulations of lattice-QCD \cite{McNeile:2006qy}, as illustrated in Fig.\,\ref{fig:fRadial}.
It is now recognised that the suppression of $f_{\pi_1}$ is a useful benchmark, which can be used to tune and validate lattice QCD techniques that try to determine the properties of excited states mesons.

\subsection{In-hadron condensates}
\label{inhadroncondensate}
For the last thirty years, \emph{condensates}; i.e., nonzero vacuum expectation values of local operators, have been used as parameters in order to correlate and estimate essentially nonperturbative strong-interaction matrix elements \cite{Colangelo:2000dp}.  They are also basic to current algebra analyses.  It is conventionally held that such quark and gluon condensates have a physical existence, which is independent of the hadrons that express QCD's asymptotically realisable degrees-of-freedom; namely, that these condensates are not merely mass-dimensioned parameters in a theoretical truncation scheme, but in fact describe measurable spacetime-independent configurations of QCD's elementary degrees-of-freedom in a hadronless ground state.  Owing to confinement, however, this view is erroneous \cite{Brodsky:2010xf}.

One may readily recapitulate upon the argument.  To begin, note that Eq.\,(\ref{fpigen}) is the exact expression in QCD for the leptonic decay constant of a pseudoscalar meson.  It is a property of the pion and, as consideration of the integral expression reveals, it can be described as the pseudovector projection of the pion's Bethe-Salpeter wavefunction onto the origin in configuration space.  Note that the product $\psi = S \Gamma S$ is called the Bethe-Salpeter wavefunction because, when a nonrelativistic limit can validly be performed, the quantity $\psi$ at fixed time becomes the quantum mechanical wavefunction for the system under consideration.  (NB.\ In the neighborhood of the chiral limit, a value for $f_{H_{0^-}}$ can be estimated via either of two approximation formulae \protect\cite{Pagels:1979hd,Cahill:1985mh,Chang:2009zb}.  These formulae both illustrate and emphasize the role of $f_{H_{0^-}}$ as an order parameter for DCSB.)

If chiral symmetry were not dynamically broken, then in the neighborhood of the chiral limit $f_{H_{0^-}} \propto \hat m$ \cite{Holl:2004fr}.  Of course, chiral symmetry is dynamically broken in QCD \cite{Bhagwat:2003vw,Bhagwat:2006tu,Bowman:2005vx} and for the ground-state pseudoscalar
\begin{equation}
\lim_{\hat m\to 0} f_{H_{0^-}}(\hat m) = f^0_{H_{0^-}} \neq 0\,.
\end{equation}
Taken together, these last two observations express the fact that $f_{H_{0^-}}$, which is an intrinsic property of the pseudoscalar meson, is a \emph{bona fide} order parameter for DCSB.  An analysis within chiral perturbation theory \cite{Bijnens:2006zp} suggests that the chiral limit value, $f^0_{H_{0^-}}$, is $\sim 5$\% below the measured value of 92.4\,MeV; and efficacious DSE studies give a 3\% chiral-limit reduction~\cite{Maris:1997tm}.

Now, Eq.\,(\ref{rhogen}) is kindred to Eq.\,(\ref{fpigen}); it is the expression in quantum field theory which describes the \emph{pseudoscalar} projection of the pseudoscalar meson's Bethe-Salpeter wavefunction onto the origin in configuration space.  It is thus truly just another type of pseudoscalar meson decay constant.  

In this connection it is therefore notable that one may rigorously define an ``in-meson'' condensate; viz.\,\cite{Maris:1997hd,Maris:1997tm}:
\begin{equation}
\label{inpiqbq}
-\langle \bar q_{f_2} q_{f_1} \rangle^\zeta_{H_{0^-}} \equiv -
f_{H_{0^-}} \langle 0 | q_{f_2} \gamma_5 q_{f_1} |H_{0^-} \rangle
= f_{H_{0^-}} \rho_{H_{0^-}}^\zeta =: \kappa_{H_{0^-}}^\zeta(\hat m)\,.
\end{equation}
Now, using Eq.\,(\ref{gtlrelE}), one finds \cite{Maris:1997hd}
\begin{equation}
\lim_{\hat m\to 0} \kappa^\zeta_{H_{0^-}}(\hat m)
=
Z_4 \, {\rm tr}_{\rm CD}\int^\Lambda \!\!\!\! \mbox{\footnotesize $\displaystyle\frac{d^4 q}{(2\pi)^4}$} S^0(q;\zeta) =  -\langle \bar q q \rangle_\zeta^0\,.
\label{qbqpiqbq0}
\end{equation}
Hence the so-called vacuum quark condensate is, in fact, the chiral-limit value of the in-meson condensate; i.e., it describes a property of the chiral-limit pseudoscalar meson.  One can therefore argue that this condensate is no more a property of the ``vacuum'' than the pseudoscalar meson's chiral-limit leptonic decay constant.  Moreover, Ref.\,\cite{Langfeld:2003ye} establishes the equivalence of all three definitions of the so-called vacuum quark condensate: a constant in the operator product expansion \cite{Lane:1974he,Politzer:1976tv}; via the Banks-Casher formula \cite{Banks:1979yr}; and the trace of the chiral-limit dressed-quark propagator.
(NB.\, In the presence of confinement it is impossible to write a valid nonperturbative definition of a single quark or gluon annihilation operator and therefore impossible to rigorously define a second quantised vacuum for QCD.  To do so would be to answer the question: What is the state that is annihilated by an operator which is unknowable?)

The chiral-limit vacuum quark condensate is therefore qualitatively equivalent to the pseudoscalar-meson leptonic decay constant in the sense that both are obtained as the chiral-limit value of well-defined gauge-invariant hadron-to-vacuum transition amplitudes that possess a spectral representation in terms of the current-quark-mass.  Thus, whereas it might sometimes be convenient to imagine otherwise, neither is essentially a constant mass-scale that fills all spacetime.  Hence, in particular, the quark condensate can be understood as a property of hadrons themselves -- a perspective also advocated in Ref.\,\protect\cite{Casher:1974xd} and now established for all hadrons \cite{Chang:2011mu} -- which is expressed, for example, in their Bethe-Salpeter or light-front wave functions.  In the latter instance, the light-front-instantaneous quark propagator appears to play a crucial role \cite{Brodsky:2010xf,Burkardt:1998dd}.

This has enormous implications for the cosmological constant.  The universe is expanding at an ever-increasing rate and theoretical physics has tried to explain this in terms of the energy of quantum processes in vacuum; viz., condensates carry energy and so, if they exist, must contribute to the universe's dark energy, which is expressed in the computed value of the cosmological constant.  The problem is that hitherto all potential sources of this so-called vacuum energy give magnitudes that far exceed the value of the cosmological constant that is empirically determined.  This has been described as ``the biggest embarrassment in theoretical physics'' \cite{Turner:2001yu}.  However, given that, in the presence of confinement, condensates do not leak from within hadrons, then there are no space-time-independent condensates permeating the universe \cite{Brodsky:2010xf,Chang:2011mu}.  This nullifies completely their contribution to the cosmological constant and reduces the mismatch between theory and observation by a factor of $10^{46}$ \cite{Brodsky:2009zd}, and possibly by far more, if technicolour-like theories are the correct paradigm for extending the Standard Model.

\subsection{Flavourless pseudoscalar mesons}
\label{flavourless}
\begin{figure}[t]

\centerline{
\includegraphics[clip,width=0.5\textwidth]{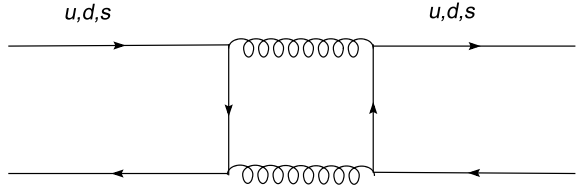}}

\caption{\label{fig:nonanomaly}
This simple flavour-mixing diagram is immaterial to the resolution of the $\eta$-$\eta^\prime$ conundrum, as is any collection of processes for which the figure may serve as a skeleton diagram.  (Straight lines denote quarks and springs denote gluons.)
}
\end{figure}

In connection with electric-charge-neutral pseudoscalar mesons, Eq.\,(\ref{mrtrelation}) is strongly modified owing to the non-Abelian anomaly.  This entails that whilst the classical action associated with QCD is invariant under $U_A(N_f)$ (non-Abelian axial transformations generated by  $\lambda_0 \gamma_5$, where $\lambda_0 \propto{\rm diag}[1,\ldots ,1_{N_f}]$), the quantum field theory is not.  The modification is particularly important to properties of $\eta$ and $\eta^\prime$ mesons.  The latter is obviously a peculiar pseudoscalar meson because its mass is far greater than that of any other light-quark pseudoscalar meson; e.g., $m_{\eta^\prime} = 1.75\, m_{\eta}$.  We note that the diagram depicted in Fig.\,\ref{fig:nonanomaly} is often cited as central to a solution of the $\eta$-$\eta^\prime$ puzzle.  However, as will become clear below, whilst it does contribute to flavour-mixing, the process is immaterial in resolving the $\eta$-$\eta^\prime$ conundrum.

The correct mass formula for flavourless pseudoscalars follows from consideration of the complete $U_A(N_f)$ Ward-Takahashi identity:
\begin{equation}
%
%
%
P_\mu \Gamma_{5\mu}^a(k;P) = {\cal S}^{-1}(k_+) i \gamma_5 {\cal F}^a
+ i \gamma_5 {\cal F}^a {\cal S}^{-1}(k_-)
- 2 i {\cal M}^{ab}\Gamma_5^b(k;P)  - {\cal A}^a(k;P)\,,
\label{avwtiG}
\end{equation}
which generalises Eq.\,(\ref{avwtimN}).  In Eq.\,(\ref{avwtiG}),
$\{{\cal F}^a | \, a=0,\ldots,N_f^2-1\}$ are the generators of $U(N_f)$ in the fundamental representation, orthonormalised according to tr${\cal F}^a {\cal F}^b= \frac{1}{2}\delta^{ab}$;
the dressed-quark propagator ${\cal S}=\,$diag$[S_u,S_d,S_s,S_c,S_b,\ldots]$ is matrix-valued;
and
\begin{equation}
{\cal M}^{ab} = {\rm tr}_F \left[ \{ {\cal F}^a , {\cal M}^\zeta \} {\cal F}^b \right],
\end{equation}
where ${\cal M}^\zeta$ is a matrix of renormalised current-quark masses and the trace is over flavour indices.

The final term in the last line of Eq.\,(\ref{avwtiG}) expresses the non-Abelian axial anomaly.  It can be written
\begin{eqnarray}
\label{amputate}
{\cal A}^a(k;P) &=&  {\cal S}^{-1}(k_+) \,\delta^{a0}\, {\cal A}_U(k;P) {\cal S}^{-1}(k_-)\,,\\
{\cal A}_U(k;P) &=& \!\!  \int\!\! d^4xd^4y\, e^{i(k_+\cdot x - k_- \cdot y)} N_f \left\langle  {\cal F}^0\,q(x)  \, {\cal Q}(0) \,   \bar q(y)
\right\rangle, \label{AU}
\end{eqnarray}
and since ${\cal A}^{a=0}(k;P)$ is a pseudoscalar, it has the general form
\begin{eqnarray}
\nonumber {\cal A}^0(k;P) &=& {\cal F}^0\gamma_5 \left[ i {\cal E}_{\cal A}(k;P) \right. \\
&& \left. + \gamma\cdot P {\cal F}_{\cal A}(k;P)  +\, \gamma\cdot k \, k\cdot P {\cal G}_{\cal A}(k;P) + \sigma_{\mu\nu} k_\mu P_\nu {\cal H}_{\cal A}(k;P)\right].
\end{eqnarray}
The matrix element in Eq.\,(\ref{AU}) represents an operator expectation value in full QCD; the operation in Eq.\,(\ref{amputate}) amputates the external quark lines; and
\begin{equation}
{\cal Q}(x) = i \frac{\alpha_s }{4 \pi} {\rm tr}_{C}\left[ \epsilon_{\mu\nu\rho\sigma} F_{\mu\nu} F_{\rho\sigma}(x)\right]  \label{topQ} = \partial_\mu K_\mu(x)
\end{equation}
is the topological charge density operator, where the trace is over colour indices and $F_{\mu\nu}=\frac{1}{2}\lambda^a F_{\mu\nu}^a$ is the matrix-valued gluon field strength tensor.  It is plain and important that only ${\cal A}^{a=0}$ is nonzero.  NB.\ While ${\cal Q}(x)$ is gauge invariant, the associated Chern-Simons current, $K_\mu$, is not.  Thus in QCD no physical state can couple to $K_\mu$ and hence no state which appears in the observable spectrum can contribute to a resolution of the so-called $U_A(1)$-problem; i.e., physical states cannot play any role in ensuring that the $\eta^\prime$ is not a Goldstone mode.

As described in Sec.\,\ref{psmassformula}, if one imagines there are $N_f$ massless quarks, then DCSB is a necessary and sufficient condition for the $a\neq 0$ components of Eq.\,(\ref{avwtiG}) to guarantee the existence of $N_f^2-1$ massless bound-states of a dressed-quark and -antiquark.  However, owing to Eq.\,(\ref{amputate}), $a=0$ in Eq.\,(\ref{avwtiG}) requires special consideration.  One case is easily covered; viz., it is clear that if ${\cal A}^{0} \equiv 0$, then the $a=0$ component of Eq.\,(\ref{avwtiG}) is no different to the others and there is an additional massless bound-state in the chiral limit.

On the other hand, the large disparity between the mass of the $\eta^\prime$-meson and the octet pseudoscalars suggests that ${\cal A}^{0} \neq 0$ in real-world QCD.  If one carefully considers that possibility, then the Goldberger-Treiman relations in Eqs.\,(\ref{gtlrelE}) -- (\ref{gtlrelH}) become \cite{Bhagwat:2007ha}
\begin{eqnarray}
\label{ewti}
2 f_{H_{0^-}}^0 E_{BS}(k;0) &= & 2 B^{0}(k^2) - {\cal E}_{\cal A}(k;0),\\
\label{fwti}
F_R^0(k;0) + 2 f_{H_{0^-}}^0 F_{BS}(k;0) & = & A^{0}(k^2) - {\cal F}_{\cal A}(k;0),\\
G_R^0(k;0) + 2 f_{H_{0^-}}^0 G_{BS}(k;0) & = & 2 \frac{d}{dk^2}A^{0}(k^2) - {\cal G}_{\cal A}(k;0),\\
\label{hwti}
H_R^0(k;0) + 2 f_{H_{0^-}}^0 H_{BS}(k;0) & = & - {\cal H}_{\cal A}(k;0),
\end{eqnarray}
It follows that the relationship
\begin{equation}
\label{calEB}
{\cal E}_{\cal A}(k;0) = 2 B^{0}(k^2) \,,
\end{equation}
is necessary and sufficient to guarantee that $\Gamma_{5\mu}^0(k;P)$, the flavourless pseudoscalar vertex, does not possess a massless pole in the chiral limit; i.e., that there are only $N_f^2-1$ massless Goldstone bosons.  Now, in the chiral limit, $B^{0}(k^2) \neq 0 $ if, and only if, chiral symmetry is dynamically broken.   Hence, the absence of an additional massless bound-state is only assured through the existence of an intimate connection between DCSB and an expectation value involving the topological charge density.

This critical connection is further highlighted by the following result, obtained through a few straightforward manipulations of Eqs.\,(\ref{avwtiG}), (\ref{amputate}) and (\ref{AU}):
\begin{eqnarray}
\langle \bar q q \rangle_\zeta^0 = - \lim_{\hat m \to 0} \kappa_{H_{0^-}}^\zeta(\hat m)
& = & -\lim_{\Lambda\to \infty}Z_4(\zeta^2,\Lambda^2)\, {\rm tr}_{\rm CD}\int^\Lambda_q\!
S^{0}(q,\zeta)  \\
& = &
\mbox{\footnotesize $\displaystyle \frac{N_f}{2}$} \int d^4 x\, \langle \bar q(x) i\gamma_5  q(x) {\cal Q}(0)\rangle^0.
\end{eqnarray}
The absence of a Goldstone boson in the $a=0$ channel is only guaranteed if this explicit identity between the chiral-limit in-meson condensate and a mixed vacuum polarisation involving the topological charge density is satisfied.

Mass formulae valid for all pseudoscalar mesons have also been obtained \cite{Bhagwat:2007ha}
\begin{equation}
\label{newmass}
%
f_{H_{0^-}}^a m_{H_{0^-}}^2 = 2\,{\cal M}^{ab} \rho_{H_{0^-}}^b + \delta^{a0} \, n_{H_{0^-}}\,,
\end{equation}
where
\begin{eqnarray}
\label{fpia} f_{H_{0^-}}^a \,  P_\mu &=& Z_2\,{\rm tr} \int^\Lambda_q
{\cal F}^a \gamma_5\gamma_\mu\, \chi_{H_{0^-}}(q;P) \,, \\
\label{cpres} i  \rho_{H_{0^-}}^a\!(\zeta)  &=& Z_4\,{\rm tr}
\int^\Lambda_q {\cal F}^a \gamma_5 \, \chi_{H_{0^-}}(q;P)\,,\\
n_{H_{0^-}} &=& \mbox{\footnotesize $\displaystyle \sqrt{\frac{N_f}{2}}$} \, \nu_{H_{0^-}} \,, \; \nu_{H_{0^-}}= \langle 0 | {\cal Q} | H_{0^-}\rangle \,.
\end{eqnarray}
For charged pseudoscalar mesons, Eq.\,(\ref{newmass}) is equivalent to Eq.\,(\ref{mrtrelation}), but the novelty of Eq.\,(\ref{newmass}) is what it expresses for flavourless pseudoscalars.  To illustrate, consider the case of a $U(N_f=3)$-symmetric mass matrix, in which all $N_f=3$ current-quark masses assume the single value $m^\zeta$, then this formula yields:
\begin{equation}
\label{etapchiral}
m_{\eta^\prime}^2 f_{\eta^\prime}^0 = n_{\eta^\prime} + 2 m^\zeta\rho_{\eta^\prime}^{0 \zeta} \,.
\end{equation}
Plainly, the $\eta^\prime$ is split from the Goldstone modes so long as $n_{\eta^\prime} \neq 0$.  Numerical simulations of lattice-QCD have confirmed this identity \protect\cite{Bardeen:2000cz,Ahmad:2005dr}.

It is important to elucidate the physical content of $n_{\eta^\prime}$.  Returning to the definition:
\begin{equation}
\nu_{\eta^\prime}= \mbox{\footnotesize $\displaystyle \sqrt{\frac{3}{2}}$} \, \langle 0 | {\cal Q} | \eta^\prime \rangle \,,
\end{equation}
it is readily seen to be another type of in-meson condensate.  It is analogous to those discussed in Sec.\,\ref{inhadroncondensate} but in this case the hadron-to-vacuum transition amplitude measures the topological content of the $\eta^\prime$.  One may therefore state that the $\eta^\prime$ is split from the Goldstone modes so long as its wavefunction possesses nonzero topological content.  This is plainly very different to requiring that the QCD vacuum is topologically nontrivial.

\begin{figure}[t]
\vspace*{-30ex}

\centerline{
\includegraphics[clip,width=0.7\textwidth]{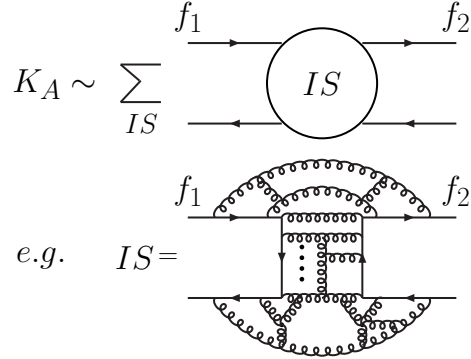}}
\vspace*{-35ex}

\caption{\label{etaglue} An illustration of the nature of the contribution to the Bethe-Salpeter kernel associated with the non-Abelian anomaly.  All terms have the ``hairpin'' structure illustrated in the lower panel.  No finite sum of such intermediate states is sufficient.  (Straight lines denote quarks, with $f_1$ and $f_2$ independent, and springs denote gluons.)
(Figure adapted from Ref.\,\protect\cite{Bhagwat:2007ha}.)}
\end{figure}

Within QCD the properties of the $\eta^\prime$ can be computed via the BSE, just like other mesons.  A nonzero value of $n_{\eta^\prime}$ can be achieved with a Bethe-Salpeter kernel that contains the contribution depicted in Fig.\,\ref{etaglue} because one may argue from Eqs.\,(\ref{AU}) and (\ref{topQ}) that an anomaly-related contribution to a meson's Bethe-Salpeter kernel cannot contain external quark or antiquark lines that are connected to the incoming lines: purely gluonic configurations must mediate, as illustrated in Fig.\,\ref{etaglue}.  Furthermore, it is straightforward to see that no finite sum of gluon exchanges can serve this purpose.  Indeed, consider any one such single contribution in the chiral limit.  It will be proportional to the total momentum and hence vanish for $P=0$, in conflict with Eq.\,(\ref{etapchiral}).  This lies behind the need for something like the Kogut-Susskind \emph{ghost}; i.e., the coupling of a massless axial-vector gauge-like field to the Chern-Simons current, which does not appear in the particle spectrum of QCD because the current is not gauge invariant.  (See Ref.\,\protect\cite{Kogut:1974kt} and Sec.\,5.1 of Ref.\,\cite{Christos:1984tu}.)

It is argued \cite{Witten:1979vv,Veneziano:1979ec} that in QCD with $N_c$ colours,
\begin{equation}
n_{\eta^\prime} \sim \frac{1}{\sqrt{N_c}}\,,
\end{equation}
and it can be seen to follow from the gap equation, the homogeneous BSE and Eqs.\,(\ref{fpia}), (\ref{cpres}) that
\begin{equation}
f_{\eta^\prime}^0 \sim \sqrt{N_c} \sim \rho_{\eta^\prime}^0(\zeta)\,.
\end{equation}
One thus obtains
\begin{equation}
m_{\eta^\prime}^2 =  \frac{n_{\eta^\prime}}{f_{\eta^\prime}^0} + 2 m(\zeta) \frac{\rho_{\eta^\prime}^0(\zeta)}{f_{\eta^\prime}^0} \,.
\end{equation}
The first term vanishes in the limit $N_c\to \infty$ while the second remains finite.  Subsequently taking the chiral limit, the $\eta^\prime$ mass approaches zero in the manner characteristic of all Goldstone modes.  (NB.\ One must take the limit $N_c\to \infty$ before the chiral limit because the procedures do not commute \cite{Narayanan:2004cp}.)  These results are realised in the effective Lagrangian of Ref.\,\cite{Di Vecchia:1979bf} in a fashion that is consistent with all the constraints of the anomalous Ward identity.  NB.\ This is not true of the so-called 't\,Hooft determinant \protect\cite{Crewther:1977ce,Crewther:1978zz,Christos:1984tu}.

Implications of the mass formula in Eq.\,(\ref{newmass}) were exemplified in Ref.\,\cite{Bhagwat:2007ha} using an elementary dynamical model that includes a one-parameter \emph{Ansatz} for that part of the Bethe-Salpeter kernel related to the non-Abelian anomaly, an illustration of which is provided in Fig.\,\ref{etaglue}.  The study compares ground-state pseudoscalar- and vector-mesons constituted from all known quarks, excluding the $t$-quark.  Amongst the notable results is a prediction for the mixing angles between neutral mesons; e.g.,
\begin{equation}
\label{valmixing}
\theta_\eta = -15.4^\circ\,,\;
\theta_{\eta^\prime} = -15.7^\circ\,.
\end{equation}
NB.\ There are necessarily two mixing angles, with each determined at the appropriate pole position in the inhomogeneous vertex.  It is interesting that the angles are approximately equal and compare well with the value inferred from a single mixing angle analysis \cite{KLEO} $\theta = -13.3^\circ \pm 1.0^\circ$.

It is worth explicating the nature of the flavour-induced difference between the $\pi^0$ and $\pi^\pm$ masses.  If one ignores mixing with mesons containing other than $u,d$-quarks; viz., works solely within $SU(N_f=2)$, then $m_{\pi^0}-m_{\pi^+}=-0.04\,$MeV.  On the other hand, the full calculation yields $m_{\pi^0}-m_{\pi^+}=-0.4\,$MeV, a factor of ten greater, and one obtains a $\pi^0$-$\eta$ mixing angle, whose value at the neutral pion mass shell is
\begin{equation}
\theta_{\pi \eta}(m_{\pi^0}^2)=1.2^\circ.
\end{equation}
For comparison, Ref.\,\cite{Green:2003qw} infers a mixing angle of $0.6^\circ \pm 0.3^\circ$ from a $K$-matrix analysis of the process $p\, d \rightarrow\, ^3$He$\,\pi^0$.  Plainly, mixing with the $\eta$-meson is the dominant non-electromagnetic contribution to the $\pi^\pm$-$\pi^0$ mass splitting.  The analogous angle at the $\eta$ mass-shell is
\begin{equation}
\theta_{\pi \eta}(m_{\eta}^2)=1.3^\circ.
\end{equation}

The angles in Eq.\,(\ref{valmixing}) correspond to
\begin{eqnarray}
\label{pi0f}
|\pi^0\rangle & \sim & 0.72 \, \bar u u - 0.69 \, \bar d d - 0.013 \, \bar s s\,, \\
\label{pi8f}
|\eta\rangle & \sim & 0.53\, \bar u u + 0.57 \, \bar d d - 0.63 \, \bar s s\,, \\
\label{pi9f}
|\eta^\prime\rangle & \sim & 0.44\, \bar u u + 0.45 \, \bar d d + 0.78 \, \bar s s \,.
\end{eqnarray}
Evidently, in the presence of a sensible amount of isospin breaking, the $\pi^0$ is still predominantly characterised by ${\cal F}^3$ but there is a small admixture of $\bar ss$.  It is found in Ref.\,\cite{Bhagwat:2007ha} that mixing with the $\pi^0$ has a similarly modest impact on the flavour content of the $\eta$ and $\eta^\prime$.  It's effect on their masses is far less.

\section{Expressing DCSB in Bound-States}
\label{sec:BSEK}
Despite the successes achieved with the systematic scheme, its practical application has numerous shortcomings.
For example, the leading-order truncation is accurate for electrically-charged ground-state pseudoscalar- and vector-mesons because corrections in these channels largely cancel, owing to parameter-free preservation of the Ward-Takahashi identities.  However, they do not cancel in other channels \cite{Roberts:1996jx,Roberts:1997vs,detmold,Bhagwat:2004hn}.  Hence studies based on the rainbow-ladder truncation, or low-order improvements thereof, have usually provided poor results for scalar- and axial-vector-mesons \cite{Cloet:2007pi,Burden:1996nh,Watson:2004kd,Maris:2006ea,Fischer:2009jm,%
Krassnigg:2009zh}, produced masses for exotic states that are too low in comparison with other estimates \cite{Cloet:2007pi,Qin:2011xq,Burden:1996nh,Krassnigg:2009zh}, and exhibit gross sensitivity to model parameters for tensor-mesons \cite{Krassnigg:2010mh} and excited states \cite{Holl:2004fr,Holl:2004un,Qin:2011xq}.  In these circumstances one must conclude that physics important to these states is omitted.
One anticipates therefore that significant qualitative advances in understanding the essence of QCD could be made with symmetry-preserving kernel \emph{Ans\"atze} that express important additional nonperturbative effects, which are impossible to capture in any finite sum of contributions.  Such an approach has recently become available \cite{Chang:2009zb} and is well worth summarising herein.

\subsection{Building the Bethe-Salpeter kernel}
\label{sec:building}
\begin{figure}[t]
\centerline{%
\includegraphics[width=0.50\textwidth]{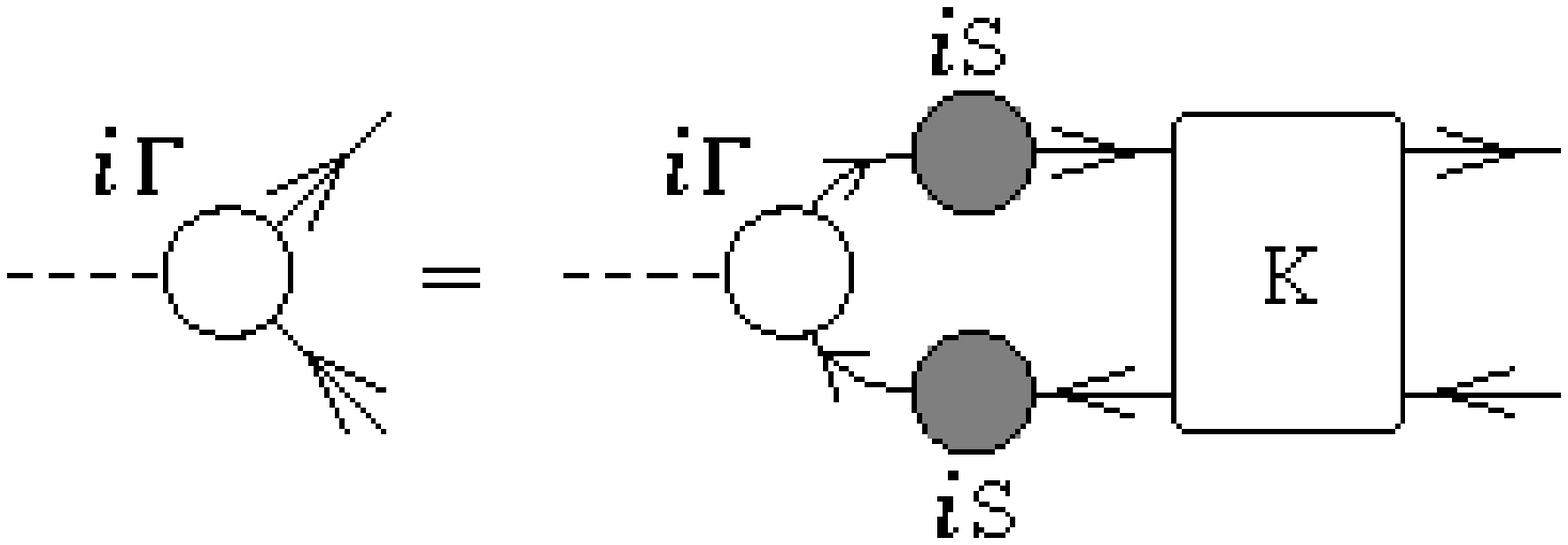}}
\centerline{%
\includegraphics[width=0.68\textwidth]{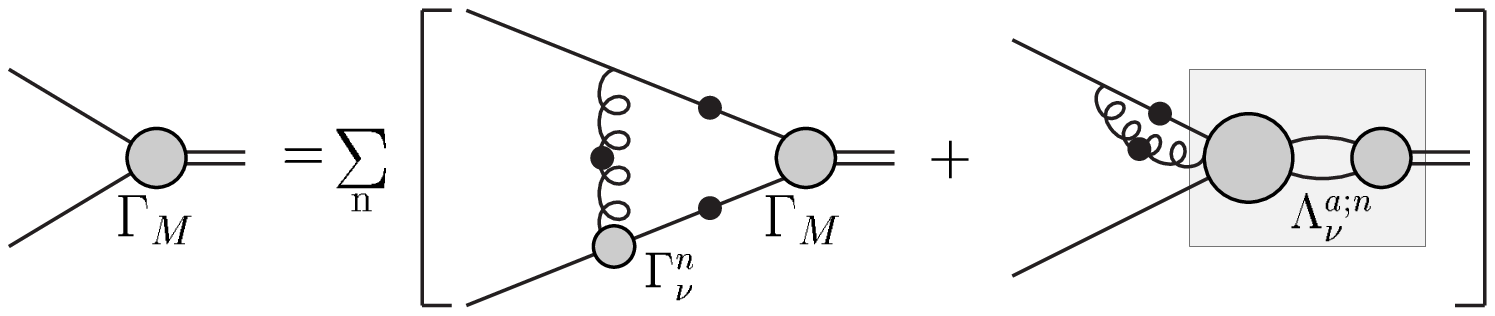}}
\caption{\label{detmoldkernel}
Omitting the inhomogeneity, the \emph{upper panel} illustrates the textbook form of the Bethe-Salpeter equation, Eq.\,(\protect\ref{bsetextbook}), whereas the \emph{lower panel} depicts the form expressed in Eq.\,(\protect\ref{genbse}).  The reversal of the total-momentum's flow is immaterial here.
NB.\ In any symmetry-preserving truncation, beyond the leading-order identified in Ref.\,\protect\cite{Bender:1996bb}; i.e., rainbow-ladder, the Bethe-Salpeter kernel is nonplanar even if the vertex in the gap equation is planar \protect\cite{detmold}.
(Figure adapted from Refs.\,\protect\cite{Roberts:1994dr,detmold}.)}
\end{figure}

Consider, e.g., flavoured pseudoscalar and axial-vector mesons, which appear as poles in the inhomogeneous BSE for the axial-vector vertex, $\Gamma_{5\mu}^{fg}$, where $f,g$ are flavour labels.  An exact form of that equation is ($k$, $q$ are relative momenta, $P$ is the total momentum flowing into the vertex, and $q_\pm = q\pm P/2$, etc.)
\begin{eqnarray}
\nonumber
\Gamma_{5\mu}^{fg}(k;P) &=& Z_2 \gamma_5\gamma_\mu - \int_q^\Lambda g^2 D_{\alpha\beta}(k-q) \frac{\lambda^a}{2}\,\gamma_\alpha S_f(q_+) \Gamma_{5\mu}^{fg}(q;P) S_g(q_-) \frac{\lambda^a}{2}\,\Gamma_\beta^g(q_-,k_-) \\
&  + &\int^\Lambda_q g^2D_{\alpha\beta}(k-q)\, \frac{\lambda^a}{2}\,\gamma_\alpha S_f(q_+) \frac{\lambda^a}{2} \Lambda_{5\mu\beta}^{fg}(k,q;P), \rule{1em}{0ex} \label{genbse}
\end{eqnarray}
where $\Lambda_{5\mu\beta}^{fg}$ is a 4-point Schwinger function.  (The pseudoscalar vertex satisfies an analogue of Eq.\,(\ref{genbse}).)  This form of the BSE was first written in Ref.\,\cite{detmold} and is illustrated in the lower-panel of Fig.\,(\ref{detmoldkernel}).  The diagrammatic content of the right-hand-side is completely equivalent to that of Eq.\,(\ref{bsetextbook}), which is depicted in the upper-panel of the figure.  However, in striking qualitative opposition to that textbook equation, Eq.\,(\ref{genbse}) partly embeds the solution vertex in the four-point function, $\Lambda$, whilst simultaneously explicating a part of the effect of the dressed-quark-gluon vertex.  This has the invaluable consequence of enabling the derivation of both an integral equation for the new Bethe-Salpeter kernel, $\Lambda$, in which the driving term is the dressed-quark-gluon vertex \cite{detmold}, and a Ward-Takahashi identity relating $\Lambda$ to that vertex \cite{Chang:2009zb}.  No similar equations have yet been found for $K$ and hence the textbook form of the BSE, whilst tidy, is very limited in its capacity to expose the effects of DCSB in hadron physics.

As emphasised above, no study of light-quark hadrons is dependable if it fails to comply with the axial-vector Ward-Takahashi identity, Eq.\,(\ref{avwtimN}).  The condition
\begin{equation}
P_\mu \Lambda_{5\mu\beta}^{fg}(k,q;P) + i [m_f(\zeta)+m_g(\zeta)] \Lambda_{5\beta}^{fg}(k,q;P)= \Gamma_\beta^f(q_+,k_+) \, i\gamma_5+ i\gamma_5 \, \Gamma_\beta^g(q_-,k_-) \,, \label{LavwtiGamma}
\end{equation}
where $\Lambda_{5\beta}^{fg}$ is the analogue of $\Lambda_{5\mu\beta}^{fg}$ in the pseudoscalar equation, is necessary and sufficient to ensure the Ward-Takahashi identity is satisfied by the solution of Eqs.\,(\ref{gendseN}) and (\ref{genbse}) \cite{Chang:2009zb}.

Consider Eq.\,(\ref{LavwtiGamma}).  Rainbow-ladder is the lead\-ing-or\-der term in the systematic DSE truncation scheme of Refs.\,\cite{Munczek:1994zz,Bender:1996bb}.  It corresponds to $\Gamma_\nu^f=\gamma_\nu$, in which case Eq.\,(\ref{LavwtiGamma}) is solved by $\Lambda_{5\mu\beta}^{fg}\equiv 0 \equiv \Lambda_{5\beta}^{fg}$.  This is the solution that indeed provides the rainbow-ladder forms of Eq.\,(\ref{genbse}).  Such consistency will be apparent in any valid systematic term-by-term improvement of the rainbow-ladder truncation.

However, Eq.\,(\ref{LavwtiGamma}) is far more than just a device for checking a truncation's consistency.  For, just as the vector Ward-Takahashi identity has long been used to build \emph{Ans\"atze} for the dressed-quark-photon vertex \cite{Roberts:1994dr,Ball:1980ay,Kizilersu:2009kg}, Eq.\,(\ref{LavwtiGamma}) provides a tool for constructing a symmetry-preserving kernel of the BSE that is matched to any reasonable \emph{Ansatz} for the dressed-quark-gluon vertex which appears in the gap equation.  With this powerful capacity, Eq.\,(\ref{LavwtiGamma}) achieves a goal that has been sought ever since the Bethe-Salpeter equation was introduced \cite{Salpeter:1951sz}.  As will become apparent, it produces a symmetry-preserving kernel that promises to enable the first reliable Poincar\'e invariant calculation of the spectrum of mesons with masses larger than 1\,GeV.

The utility of Eq.\,(\ref{LavwtiGamma}) was illustrated in Ref.\,\cite{Chang:2009zb} through an application to ground state pseudoscalar and scalar mesons composed of equal-mass $u$- and $d$-quarks.  To this end, it was supposed that in Eq.\,(\ref{gendseN}) one employs an \emph{Ansatz} for the quark-gluon vertex which satisfies
\begin{equation}
P_\mu i \Gamma_\mu^f(k_+,k_-) = {\cal B}(P^2)\left[ S_f^{-1}(k_+) - S_f^{-1}(k_-)\right] , \label{wtiAnsatz}
\end{equation}
with ${\cal B}$ flavour-independent.  (NB.\ While the true quark-gluon vertex does not satisfy this identity, owing to the form of the Slavnov-Taylor identity which it does satisfy, it is plausible that a solution of Eq.\,(\protect\ref{wtiAnsatz}) can provide a reasonable pointwise approximation to the true vertex.)  Given Eq.\,(\ref{wtiAnsatz}), then Eq.\,(\ref{LavwtiGamma}) entails ($l=q-k$)
\begin{equation}
i l_\beta \Lambda_{5\beta}^{fg}(k,q;P) =
{\cal B}(l)^2\left[ \Gamma_{5}^{fg}(q;P) - \Gamma_{5}^{fg}(k;P)\right], \label{L5beta}
\end{equation}
with an analogous equation for $P_\mu l_\beta i\Lambda_{5\mu\beta}^{fg}(k,q;P)$.  This identity can be solved to obtain
\begin{equation}
\Lambda_{5\beta}^{fg}(k,q;P)  :=  {\cal B}((k-q)^2)\, \gamma_5\,\overline{ \Lambda}_{\beta}^{fg}(k,q;P) \,, \label{AnsatzL5a}
\end{equation}
with, using an obvious analogue of Eq.\,(\ref{genGpi}),
\begin{eqnarray}
\nonumber
\lefteqn{\overline{ \Lambda}_{\beta}^{fg}(k,q;P) =
2 \ell_\beta \, [ i \Delta_{E_5}(q,k;P)+ \gamma\cdot P \Delta_{F_5}(q,k;P)] + \gamma_\beta \, \Sigma_{G_5}(q,k;P) }\\
&+&    2 \ell_\beta \,  \gamma\cdot\ell\, \Delta_{G_5}(q,k;P)+[ \gamma_\beta,\gamma\cdot P]
\Sigma_{H_5}(q,k;P) + 2 \ell_\beta  [ \gamma\cdot\ell ,\gamma\cdot P]  \Delta_{H_5}(q,k;P) \,,\label{AnsatzL5b}
\end{eqnarray}
where $\ell=(q+k)/2$, $\Sigma_{\Phi}(q,k;P) = [\Phi(q;P)+\Phi(k;P)]/2$ and $\Delta_{\Phi}(q,k;P) = [\Phi(q;P)-\Phi(k;P)]/[q^2-k^2]$.

Now, given any \emph{Ansatz} for the quark-gluon vertex that satisfies Eq.\,(\ref{wtiAnsatz}), then the pseudoscalar analogue of Eq.\,(\ref{genbse}), and Eqs.\,(\ref{gendseN}), (\ref{AnsatzL5a}), (\ref{AnsatzL5b}) provide a symmetry-preserving closed system whose solution predicts the properties of pseudoscalar mesons.
The relevant scalar meson equations are readily derived.  (NB.\ The role played by resonant contributions to the kernel in the scalar channel \protect\cite{Holl:2005st,Pelaez:2006nj,RuizdeElvira:2010cs} is not being overlooked but they are not pertinent here.  Further comments appear in Sec.\,\protect\ref{sec:a1rho}.)
With these systems one can anticipate, elucidate and understand the influence on hadron properties of the rich nonperturbative structure expected of the fully-dressed quark-gluon vertex in QCD: in particular, that of the dynamically generated dressed-quark mass function, whose impact is quashed at any finite order in the truncation scheme of Ref.\,\cite{Bender:1996bb}.

To proceed one need only specify the gap equation's kernel because, as noted above, the BSEs are completely defined therefrom.  To complete the illustration \cite{Chang:2009zb} a simplified form of the effective interaction in Ref.\,\cite{Maris:1997tm} was employed and two vertex \emph{Ans\"atze} were compared; viz., the bare vertex $\Gamma_\mu^g = \gamma_\mu$, which defines the rainbow-ladder truncation of the DSEs and omits vertex dressing; and the Ball-Chiu (BC) vertex \cite{Ball:1980ay}, which nonperturbatively incorporates vertex dressing associated with DCSB:
\begin{equation}
i\Gamma^g_\mu(q,k)  =
i\Sigma_{A^g}(q^2,k^2)\,\gamma_\mu+
2 \ell_\mu \left[i\gamma\cdot \ell \,
\Delta_{A^g}(q^2,k^2) + \Delta_{B^g}(q^2,k^2)\right] \!.
\label{bcvtx}
\end{equation}

\begin{figure}[t]
\vspace*{-5ex}

\begin{minipage}[t]{\textwidth}
\begin{minipage}[t]{0.49\textwidth}
\leftline{\includegraphics[clip,width=1.0\textwidth]{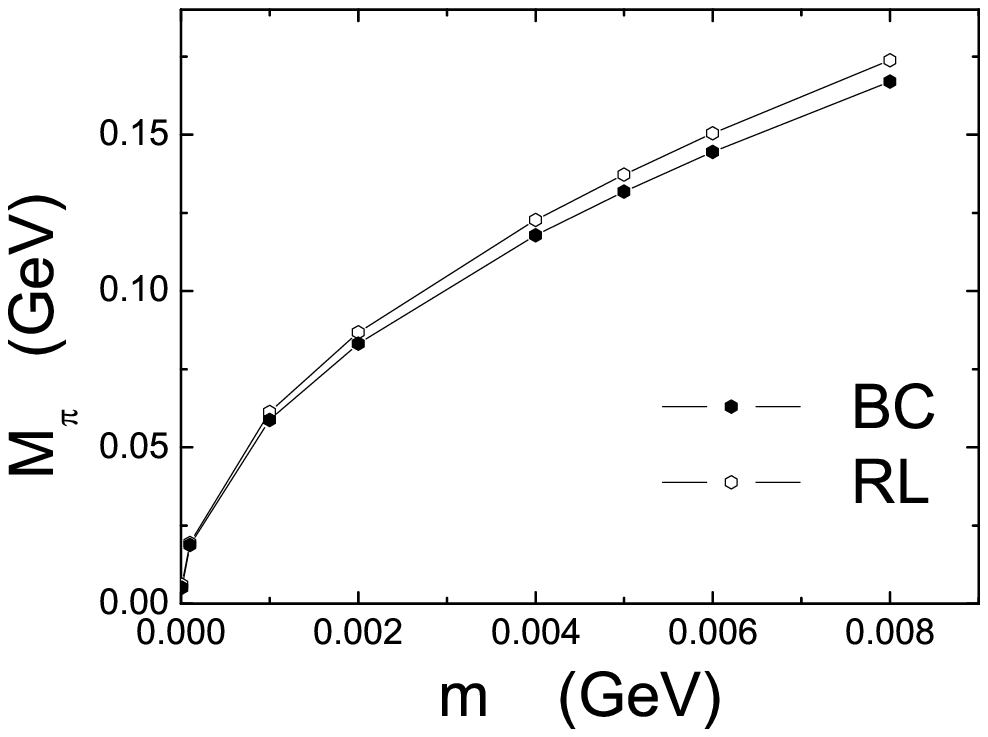}}
\end{minipage}
\begin{minipage}[t]{0.49\textwidth}
\rightline{\includegraphics[clip,width=1.0\textwidth]{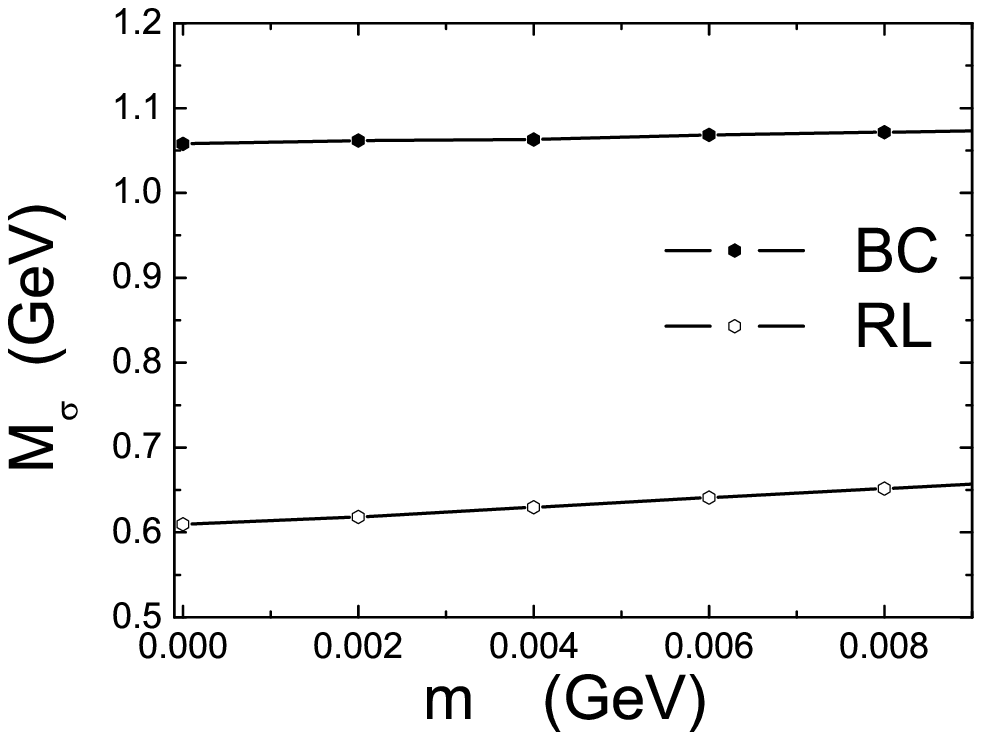}}
\end{minipage}\vspace*{3ex}
\end{minipage}
\vspace*{-4ex}



\caption{\label{massDlarge} Dependence of pseudoscalar (left panel) and scalar (right) meson masses on the current-quark mass, $m$.  The Ball-Chiu vertex (BC) result is compared with the rainbow-ladder (RL) result.  (Figure adapted from Ref.\,\protect\cite{Chang:2009zb}.)}
\end{figure}

A particular novelty of the study is that one can calculate the current-quark-mass-dependence of meson masses using a symmetry-preserving DSE truncation whose diagrammatic content is unknown. That dependence is depicted in Fig.\,\ref{massDlarge} and compared with the rainbow-ladder result.  The $m$-dependence of the pseudoscalar meson's mass provides numerical confirmation of the algebraic fact that the axial-vector Ward-Takahashi identity is preserved by both the rainbow-ladder truncation and the BC-consistent \emph{Ansatz} for the Bethe-Salpeter kernel.  The figure also shows that the axial-vector Ward-Takahashi identity and DCSB conspire to shield the pion's mass from material variation in response to dressing the quark-gluon vertex \cite{Roberts:2007jh,detmold,Bhagwat:2004hn}.

As noted in Ref.\,\cite{Chang:2009zb}, a rainbow-ladder kernel with realistic interaction strength yields
\begin{equation}
\label{epsilonRL}
\varepsilon_\sigma^{\rm RL} := \left.\frac{2 M(0) - m_\sigma }{2 M(0)}\right|_{\rm RL} = (0.3 \pm 0.1)\,,
\end{equation}
which can be contrasted with the value obtained using the BC-consistent Bethe-Salpeter kernel; viz.,
\begin{equation}
\label{epsilonBC}
\varepsilon_\sigma^{\rm BC} \lsim 0.1\,.
\end{equation}
Plainly, significant additional repulsion is present in the BC-consistent truncation of the scalar BSE.

Scalar mesons are commonly identified as $^3\!P_0$ states.  This assignment expresses a constituent-quark-model perspective, from which a $J^{PC}=0^{++}$ fermion-antifermion bound-state must have the constituents' spins aligned and one unit of constituent orbital angular momentum.  Hence a scalar is a spin and orbital excitation of a pseudoscalar meson.  Of course, no constituent-quark-model can be connected systematically with QCD.  Nevertheless, as observed in Sec.\,\ref{sect:HP}, the presence of orbital angular momentum in a hadron's rest frame is a necessary consequence of Poincar\'e covariance and the momentum-dependent vector-boson-exchange character of QCD \cite{Roberts:2007ji,Bhagwat:2006xi,Bhagwat:2006pu}, so there is a realisation in QCD of the quark-model anticipation.

Extant studies of realistic corrections to the rainbow-ladder truncation show that they reduce hyperfine splitting \cite{Bhagwat:2004hn}.  Hence, with the comparison between Eqs.\,(\ref{epsilonRL}) and (\ref{epsilonBC}) one has a clear indication that in a Poincar\'e covariant treatment the BC-consistent truncation magnifies spin-orbit splitting, an effect which can be attributed to the influence of the quark's dynamically-enhanced scalar self-energy \cite{Roberts:2007ji} in the Bethe-Salpeter kernel.

\subsection{Quark anomalous chromomagnetic moment}
\label{spectrum2}
It was conjectured in Ref.\,\cite{Chang:2009zb} that a full realisation of DCSB in the Bethe-Salpeter kernel will have a big impact on mesons with mass greater than 1\,GeV.  Moreover, that it can overcome a longstanding failure of theoretical hadron physics.  Namely, no extant continuum hadron spectrum calculation is believable because all symmetry preserving studies produce a splitting between vector and axial-vector mesons that is far too small: just one-quarter of the experimental value (see, e.g., Refs.\,\cite{Watson:2004kd,Maris:2006ea,Fischer:2009jm}).  Following that conjecture, there have been significant developments, which will now be related \cite{Chang:2010hb}.

In Dirac's relativistic quantum mechanics, a fermion with charge $q$ and mass $m$, interacting with an electromagnetic field, has a magnetic moment $\mu= q/[2 m]$.  This prediction held true for the electron until improvements in experimental techniques enabled the discovery of a small deviation \cite{Foley:1948zz}, with the moment increased by a multiplicative factor: $1.00119\pm 0.00005$.  This correction was explained by the first systematic computation using renormalised quantum electrodynamics (QED) \cite{Schwinger:1948iu}:
\begin{equation}
\label{anommme}
\frac{q}{2m} \to \left(1 + \frac{\alpha}{2\pi}\right) \frac{q}{2m}\,,
\end{equation}
where $\alpha$ is QED's fine structure constant.  The agreement with experiment established quantum electrodynamics as a valid tool.  The correction defines the electron's \emph{anomalous magnetic moment}, which is now known with extraordinary precision and agrees with theory at O$(\alpha^5)$ \cite{Mohr:2008fa}.

The fermion-photon coupling in QED is described by:
\begin{equation}
\label{QEDinteraction}
\int d^4\! x\, i q \,\bar\psi(x) \gamma_\mu \psi(x)\,A_\mu(x)\,,
\end{equation}
where $\psi(x)$, $\bar\psi(x)$ describe the fermion field and $A_\mu(x)$ describes the photon.  This interaction generates the following electromagnetic current for an on-shell Dirac fermion ($k=p_f -p_i$),
\begin{equation}
\label{ecurrent}
i q \, \bar u(p_f) \left[ \gamma_\mu F_1(k^2)+ \frac{1}{2 m} \sigma_{\mu\nu} k_\nu F_2(k^2)\right] u(p_i)\,,
\end{equation}
where: $F_1(k^2)$, $F_2(k^2)$ are form factors; and $u(p)$, $\bar u(p)$ are electron spinors.  Using their Euclidean space definition, one can derive a Gordon-identity; viz., with $2 \ell=p_f + p_i$,
\begin{equation}
\label{Gordon}
2 m \, \bar u(p_f) i \gamma_\mu u(p_i) = \bar u(p_f)\left[ 2 \ell_\mu + i \sigma_{\mu\nu} k_\nu \right]u(p_i)\,.
\end{equation}
With this rearrangement one sees that for massive fermions the interaction can be decomposed into two terms: the first describes the spin-independent part of the fermion-photon interaction, and is common to spin-zero and spin-half particles, whilst the second expresses the spin-dependent, helicity flipping part.
Moreover, one reads from Eqs.\,(\ref{ecurrent}) and (\ref{Gordon}) that a point-particle in the absence of radiative corrections has $F_1 \equiv 1$ and $F_2 \equiv 0$, and hence Dirac's value for the magnetic moment.  The anomalous magnetic moment in Eq.\,(\ref{anommme}) corresponds to $F_2(0) = \alpha/2\pi$.

One infers from Eq.\,(\ref{Gordon}) that an anomalous contribution to the magnetic moment can be associated with an additional interaction term:
\begin{equation}
\label{anominteraction}
\int d^4\! x\, \mbox{\small $\frac{1}{2}$} q \, \bar \psi(x) \sigma_{\mu\nu} \psi(x) F_{\mu\nu}(x)\,,
\end{equation}
where $F_{\mu\nu}(x)$ is the gauge-boson field strength tensor.  This term is invariant under local $U(1)$ gauge transformations but is not generated by minimal substitution in the action for a free Dirac field.

Consider the effect of the global chiral transformation $\psi(x) \to \exp(i \theta\gamma_5) \psi(x)$.  The term in Eq.\,(\ref{QEDinteraction}) is invariant.  However, the interaction of Eq.\,(\ref{anominteraction}) is not.  These observations facilitate the understanding of a general result: $F_2\equiv 0$ for a massless fermion in a quantum field theory with chiral symmetry realized in the Wigner mode; i.e., when the symmetry is not dynamically broken.  A firmer conclusion can be drawn.  For $m=0$ it follows from Eq.\,(\ref{Gordon}) that Eq.\,(\ref{QEDinteraction}) does not mix with the helicity-flipping interaction of Eq.\,(\ref{anominteraction}) and hence a massless fermion does not possess a measurable magnetic moment.

A reconsideration of Ref.\,\cite{Schwinger:1948iu} reveals no manifest conflict with these facts.  The perturbative expression for $F_2(0)$ contains a multiplicative numerator factor of $m$ and the usual analysis of the denominator involves steps that are only valid for $m\neq 0$.  Fundamentally, there is no conundrum because QED is not an asymptotically free theory and hence does not have a well-defined nonperturbative chiral limit.

On the other hand, in QCD the chiral limit is rigorously defined nonperturbatively \cite{Maris:1997tm}; and the analogue of Schwinger's one-loop calculation can be carried out to find an anomalous \emph{chromo}-magnetic moment for the quark.  There are two diagrams in this case: one similar in form to that in QED; and another owing to the gluon self-interaction.  One reads from Ref.\,\cite{Davydychev:2000rt} that the perturbative result vanishes in the chiral limit.  However, Fig.\,\ref{gluoncloud} demonstrates that chiral symmetry is dynamically broken in QCD and one must therefore ask whether this affects the chromomagnetic moment.

Of course, it does, and it is now known that this is signalled by the appearance of $\Delta_{B^g}$ in Eq.\,(\ref{bcvtx}).  Contemporary simulations of lattice-regularized QCD \cite{Skullerud:2003qu} and DSE studies \cite{Bhagwat:2004kj} agree that
\begin{equation}
\label{alpha3B}
\lambda_3(p,p;0) \approx \frac{d}{dp^2} B(p^2,\zeta)
\end{equation}
and also on the form of $\lambda_1(p,p;0)$, which is functionally similar to $A(p^2,\zeta)$.  However, owing to non-orthogonality of the tensors accompanying $\lambda_1$ and $\lambda_2$, it is difficult to obtain a lattice signal for $\lambda_2$.  We therefore consider the DSE prediction for $\lambda_2$ in Ref.\,\cite{Bhagwat:2004kj} more reliable.

As pointed out above, perturbative massless-QCD conserves helicity so the quark-gluon vertex cannot perturbatively have a term with the helicity-flipping characteristics of $\lambda_3$.  Equation~(\ref{alpha3B}) is thus remarkable, showing that the dressed-quark-gluon vertex contains at least one chirally-asymmetric component whose origin and size owe solely to DCSB; and Sec.\,\ref{sec:building} illustrates that $\lambda_3$ has a material impact on the hadron spectrum.

This reasoning is extended in Ref.\,\cite{Chang:2010hb}: massless fermions in gauge field theories cannot possess an anomalous chromo/electro-magnetic moment because the term that describes it couples left- and right-handed fermions; however, if chiral symmetry is strongly broken dynamically, then the fermions should also posses large anomalous magnetic moments.  Such an effect is expressed in the dressed-quark-gluon vertex via a term
\begin{equation}
\label{qcdanom1}
\Gamma_\mu^{\rm acm_5} (p_f,p_i;k) = \sigma_{\mu\nu} k_\nu \, \tau_5(p_f,p_i,k)\,.
\end{equation}

That QCD generates a strongly momentum-dependent chromomagnetic form factor in the quark-gluon vertex, $\tau_5$, with a large DCSB-component, is confirmed in Ref.\,\cite{Skullerud:2003qu}.  Only a particular kinematic arrangement was readily accessible in that lattice simulation but this is enough to learn that, at the current-quark mass considered: $\tau_5$ is roughly two orders-of-magnitude larger than the perturbative form; and
\begin{equation}
\label{boundtau5}
\forall p^2>0: \; |\tau_5(p,-p;2 p)| \gtrsim |\lambda_3(p,p;0)|\,.
\end{equation}
The magnitude of the lattice result is consistent with instanton-liquid model estimates \cite{Kochelev:1996pv,Diakonov:2002fq}.

This large chromomagnetic moment is likely to have a broad impact on the properties of light-quark systems \cite{Diakonov:2002fq,Ebert:2005es}.  In particular, as will be illustrated in Sec.\,\ref{sec:a1rho}, it can explain the longstanding puzzle of the mass splitting between the $a_1$- and $\rho$-mesons in the hadron spectrum \cite{Chang:2011ei}.  Here a different novel effect will be elucidated; viz., the manner in which the quark's chromomagnetic moment generates a quark anomalous \emph{electro}magnetic moment.  This demonstration is only possible now that the method of Ref.\,\cite{Chang:2009zb} is available.  It was accomplished \cite{Chang:2010hb} using the same simplification of the effective interaction in Ref.\,\cite{Maris:1997tm} that produced Figs.\,\ref{massDlarge}.

In order to understand the vertex \emph{Ansatz} used in Ref.\,\cite{Chang:2010hb}, it is necessary to return to perturbation theory.  One can determine from Ref.\,\cite{Davydychev:2000rt} that at leading-order in the coupling, $\alpha_s$, the three-gluon vertex does not contribute to the QCD analogue of Eq.\,(\ref{anommme}) and the Abelian-like diagram produces the finite and negative correction $(-\alpha_s/[12 \pi])$.
The complete cancellation of ultraviolet divergences occurs only because of the dynamical generation of another term in the transverse part of the quark-gluon vertex; namely,
\begin{equation}
\Gamma_\mu^{\rm acm_4}(p_f,p_i) = [ \ell_\mu^{\rm T} \gamma\cdot  k + i \gamma_\mu^{\rm T} \sigma_{\nu\rho}\ell_\nu k_\rho] \tau_4(p_f,p_i)\,,
\end{equation}
with $T_{\mu\nu} = \delta_{\mu\nu} - k_\mu k_\nu/k^2$, $a_\mu^{\rm T} := T_{\mu\nu}a_\nu$.

Cognisant of this, one may use a simple \emph{Ansatz} to express the dynamical generation of an anomalous chromomagnetic moment via the dressed-quark gluon vertex; viz.,
\begin{eqnarray}
\label{ourvtx}
\tilde\Gamma_\mu(p_f,p_i)  & = & \Gamma_\mu^{\rm BC}(p_f,p_i) +
\Gamma_\mu^{\rm acm}(p_f,p_i)\,,\\
\Gamma_\mu^{\rm acm}(p_f,p_i) &=& \Gamma_\mu^{\rm acm_4}(p_f,p_i) + \Gamma_\mu^{\rm acm_5}(p_f,p_i)\,,
\end{eqnarray}
with $\tau_5(p_f,p_i) =  (-7/4)\Delta_B(p_f^2,p_i^2)$, as discussed above, and
\begin{equation}
\tau_4(p_f,p_i) = {\cal F}(z) \bigg[  \frac{1-2\eta}{M_E}\Delta_B(p_f^2,p_i^2) - \Delta_A(p_f^2,p_i^2) \bigg]. \label{tau4}
\end{equation}
The damping factor ${\cal F}(z)=(1- \exp(-z))/z$, $z=(p_i^2 + p_f^2- 2 M_E^2)/\Lambda_{\cal F}^2$, $\Lambda_{\cal F}=1\,$GeV, simplifies numerical analysis; and $M_E=\{ s| s>0, s = M^2(s)\}$ is the Euclidean constituent-quark mass.

A confined quark does not possess a mass-shell (Sec.\,\ref{Sect:Conf}).  Hence, one cannot unambiguously assign a single value to its anomalous magnetic moment.  One can nonetheless compute a magnetic moment distribution.  At each value of $p^2$, spinors can be defined to satisfy the free-particle Euclidean Dirac equation with mass $m\to M(p^2)=:\varsigma$, so that
\begin{equation}
\bar u(p_f;\varsigma) \, \Gamma_\mu( p_f,p_i;k)\,  u(p_i;\varsigma)
= \bar u(p_f) [ F_1(k^2) \gamma_\mu + \frac{1}{2 \varsigma} \,\sigma_{\mu \nu} k_\nu F_2(k^2)] u(p_i)
\label{GenSpinors}
\end{equation}
and then, from Eqs.\,(\ref{ourvtx}) -- (\ref{tau4}),
\begin{equation}
\label{kappaacm}
\kappa^{\rm acm}(\varsigma) = \frac{ - 2 \varsigma \, \eta \delta_B^{\varsigma}}
    {\sigma_A^{\varsigma} - 2 \varsigma^2 \delta_A^{\varsigma}+ 2 \varsigma \delta_B^{\varsigma} }\,,
\end{equation}
where $\sigma_A^{\varsigma} = \Sigma_A(\varsigma,\varsigma)$, $\delta_A^{\varsigma} = \Delta_A(\varsigma,\varsigma)$, etc.  The numerator's simplicity owes to a premeditated cancellation between $\tau_4$ and $\tau_5$, which replicates the one at leading-order in perturbation theory.
Where a comparison of terms is possible, this vertex \emph{Ansatz} is semi-quantitatively in agreement with Refs.\,\cite{Skullerud:2003qu,Bhagwat:2004kj}.  However, the presence and understanding of the role of $\Gamma_\mu^{\rm acm_4}$ is a novel contribution by Ref.\,\cite{Chang:2010hb}.  NB.\ It is apparent from Eq.\,(\ref{kappaacm}) that $\kappa^{\rm acm} \propto m^2$ in the absence of DCSB, so that $\kappa^{\rm acm}/[2m]\to 0$ in the chiral limit.

The BSE for the quark-photon vertex can be written following the method of Ref.\,\cite{Chang:2009zb}.  Since the method guarantees preservation of the Ward-Takahashi identities, the general form of the solution is
\begin{eqnarray}
\Gamma_\mu^\gamma(p_f,p_i) & = & \Gamma_\mu^{\rm BC}(p_f,p_i) + \Gamma_\mu^{\rm T}(p_f,p_i)\,,\\
\nonumber
\Gamma_\mu^{\rm T}(p_f,p_i) & = &
\gamma_\mu^{\rm T} \hat F_1
+ \sigma_{\mu\nu} k_\nu \hat F_2
+ T_{\mu\rho} \sigma_{\rho\nu} \ell_\nu \,\ell\cdot k\, \hat F_3
+ [ \ell_\mu^{\rm T} \gamma\cdot  k + i \gamma_\mu^{\rm T} \sigma_{\nu\rho}\ell_\nu k_\rho] \hat F_4
\\
 & & \,- i \ell_\mu^{\rm T} \hat F_5+ \ell_\mu^{\rm T} \gamma\cdot k \, \ell \cdot  k\, \hat F_6 - \ell_\mu^{\rm T} \gamma\cdot \ell \, \hat F_7
%
+ \ell_\mu^T \sigma_{\nu\rho} \ell_\nu k_\rho \hat F_8\,,
\end{eqnarray}
where $\{\hat F_i|i=1,\ldots,8\}$ are scalar functions of Lorentz-invariants constructed from $p_f$, $p_i$, $k$.  The Ward-Takahashi identity is plainly satisfied; viz., 
\begin{equation}
k_\mu i \Gamma_\mu(p_f,p_i) 
= S^{-1}(p_f) - S^{-1}(p_i)\,.
\end{equation}

\begin{figure}[t]
\vspace*{-5ex}

\begin{minipage}[t]{\textwidth}
\begin{minipage}[t]{0.45\textwidth}
\leftline{\includegraphics[clip,width=1.0\textwidth]{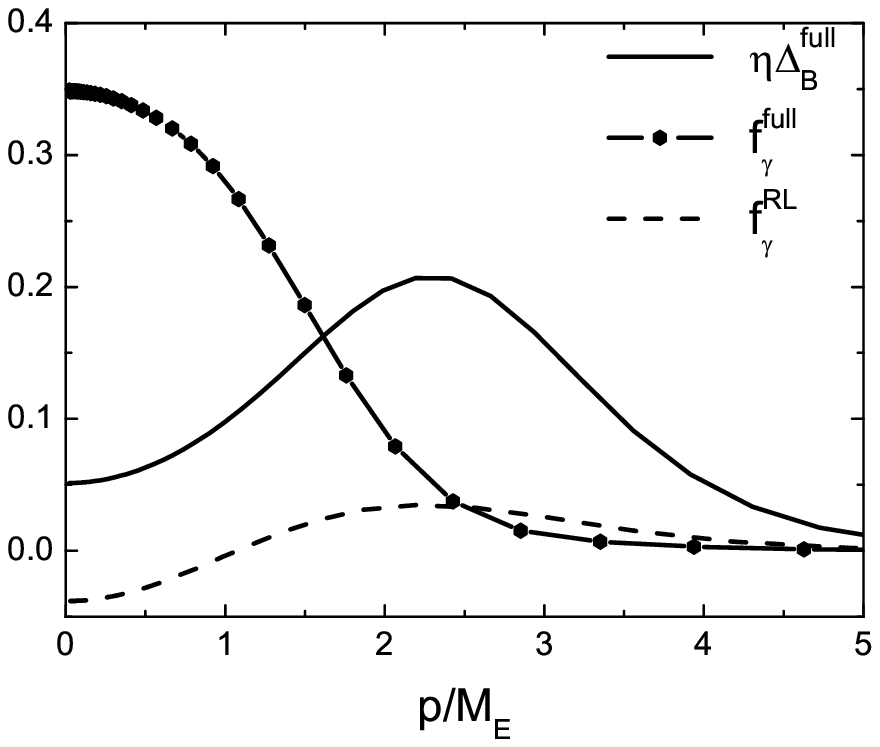}}
\end{minipage}
\begin{minipage}[t]{0.45\textwidth}
\rightline{\includegraphics[clip,width=1.05\textwidth]{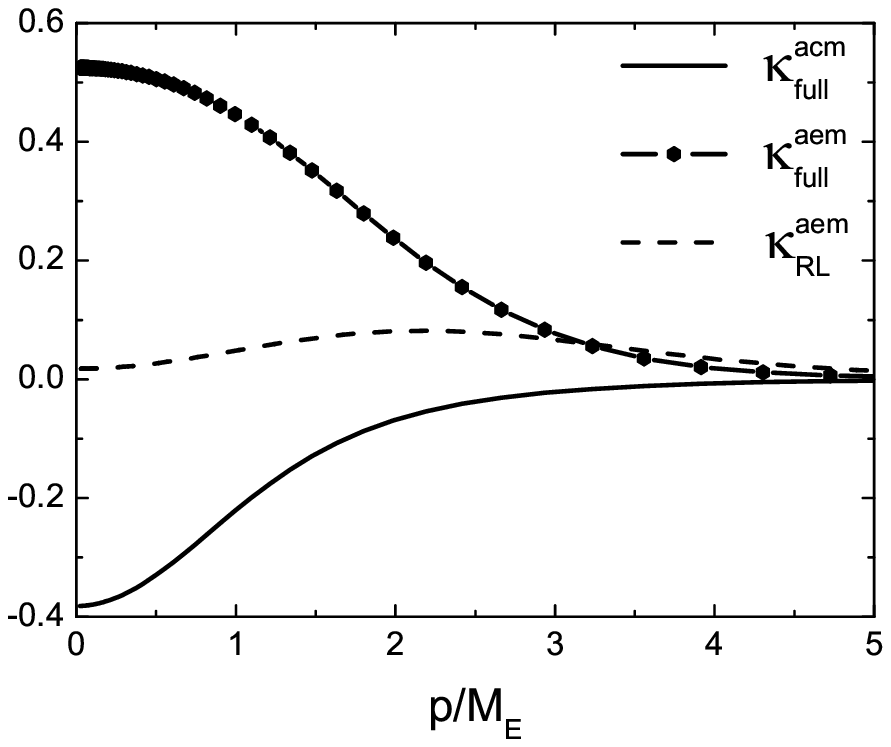}}
\end{minipage}
\end{minipage}



\caption{\label{figACM}
\emph{Left panel} -- $f_\gamma$ (GeV$^{-1}$) in Eq.\,(\protect\ref{fgamma})
cf.\ $(-7/4)\Delta_B(p^2,p^2)$, both computed using Eq.\,(\protect\ref{ourvtx}) and the same simplification of the interaction in Ref.\,\protect\cite{Maris:1997tm}.
\emph{Right panel} -- Anomalous chromo- and electro-magnetic moment distributions for a dressed-quark, computed using Eq.\,(\protect\ref{kappavalue}).
The dashed-curve in both panels is the rainbow-ladder (RL) truncation result.
(Figure adapted from Ref.\,\protect\cite{Chang:2010hb}.)}
\end{figure}

Figure~\ref{figACM} depicts the results obtained for the quark's anomalous electromagnetic moment form factor
\begin{equation}
f_\gamma(p) :=  \lim_{p_f\to p}\frac{-1}{12\,k^2}{\rm tr}\, \sigma_{\mu\nu} k_\mu \Gamma_\nu^\gamma(p_f,p)
=  \hat F_2+ \frac{1}{3}p^2 \hat F_8\,. \label{fgamma}
\end{equation}
The result is evidently sizable.
It is worth reiterating that $f_\gamma$ is completely nonperturbative: in the chiral limit, at any finite order in perturbation theory, $f_\gamma\equiv 0$.  For contrast the figure also displays the result computed in the rainbow-ladder truncation of QCD's DSEs. As the leading-order in a systematic but stepwise symmetry-preserving scheme \cite{Bender:1996bb}, this truncation only partially expresses DCSB: it is exhibited by the dressed-quark propagator but not present in the quark-gluon vertex.  In this case $f_\gamma$ is nonzero but small.  These are artefacts of the truncation that cannot be remedied at any finite order of the procedure in Ref.\,\cite{Bender:1996bb} or a kindred scheme.

Employing Eq.\,(\ref{GenSpinors}), in connection with the dressed-quark-photon vertex,
one can write an expression for the quark's anomalous electromagnetic moment distribution
\begin{equation}
\label{kappavalue}
\kappa(\varsigma)=\frac{2 \varsigma \hat F_{2} + 2 \varsigma^2 \hat F_4  +\Lambda_{\kappa}(\varsigma)}
{\sigma_{A}^{\varsigma} + \hat F_{1}-\Lambda_{\kappa}(\varsigma)}\,,
\end{equation}
where: $\Lambda_{\kappa}(\varsigma)= 2\varsigma^{2}\delta_{A}^\varsigma-2 \varsigma \delta_{B}^\varsigma -\varsigma \hat F_5 - \varsigma^2 \hat F_7$; and the $\hat F_i$ are evaluated at $p_f^2=p_i^2=M(p_f^2)^2=:\varsigma^2$, $k^2=0$.  Plainly, $\kappa(\varsigma)\equiv 0$ in the chiral limit when chiral symmetry is not dynamically broken.  Moreover, as a consequence of asymptotic freedom, $\kappa(\varsigma) \to 0$ rapidly with increasing momentum.
The computed distribution is depicted in Fig.\,\ref{figACM}.  It yields Euclidean mass-shell values:
\begin{equation}
\begin{array}{llll}
& M_{\rm full}^E = 0.44\,{\rm GeV},& \kappa_{\rm full}^{\rm acm}= -0.22\,, &
\kappa_{\rm full}^{\rm aem}= 0.45\,\\[1ex]
{\rm cf}. & M_{\rm RL}^E = 0.35\,{\rm GeV}, & \kappa_{\rm RL}^{\rm acm}= 0\,, & \kappa_{\rm RL}^{\rm aem}= 0.048 .
\end{array}
\end{equation}

It is thus apparent that DCSB produces a dressed light-quark with a momentum-dependent anomalous chromomagnetic moment, which is large at infrared momenta.  Significant amongst the consequences is the generation of an anomalous electromagnetic moment for the dressed light-quark with commensurate size but opposite sign.  (NB.\ This result was anticipated in Ref.\,\protect\cite{Bicudo:1998qb}, which argued that DCSB usually triggers the generation of a measurable anomalous magnetic moment for light-quarks.)
The infrared dimension of both moments is determined by the Euclidean constituent-quark mass.  This is two orders-of-magnitude greater than the physical light-quark current-mass, which sets the scale of the perturbative result for both these quantities.

There are two more notable features; namely, the rainbow-ladder truncation, and low-order stepwise improvements thereof, underestimate these effects by an order of magnitude; and both the $\tau_4$ and $\tau_5$ terms in the dressed-quark-gluon vertex are indispensable for a realistic description of hadron phenomena.  Whilst a simple interaction was used to illustrate these outcomes, they are robust.

These results are stimulating a reanalysis of hadron elastic and transition electromagnetic form factors, and the hadron spectrum, results of which will be described below.
Furthermore, given the magnitude of the muon ``$g_\mu-2$ anomaly'' and its assumed importance as an harbinger of physics beyond the Standard Model \cite{Jegerlehner:2009ry}, it might also be worthwhile to make a quantitative estimate of the contribution to $g_\mu-2$ from the quark's DCSB-induced anomalous moments following, e.g., the computational pattern for the hadronic light-by-light scattering component of the photon polarization tensor indicated in Ref.\,\cite{Goecke:2010if}.

\subsection{\mbox{\boldmath $a_1$}-\mbox{\boldmath $\rho$} mass splitting}
\label{sec:a1rho}
The analysis in Ref.\,\cite{Chang:2009zb} enables one to construct a symmetry-preserving kernel for the BSE given any form for $\Gamma_\mu$.  Owing to the importance of symmetries in forming the spectrum of a quantum field theory, this is a pivotal advance.  One may now use all information available, from any reliable source, to construct the best possible vertex \emph{Ansatz}.  The last section illustrated that this enables one to incorporate crucial nonperturbative effects, which any finite sum of contributions is incapable of capturing, and thereby prove that DCSB generates material, momentum-dependent anomalous chromo- and electro-magnetic moments for dressed light-quarks.

The vertex described in Sec.\,\ref{spectrum2} contains a great deal of information about DCSB.  It is the best motivated \emph{Ansatz} to date and may be used in the calculation of the masses of ground-state spin-zero and -one light-quark mesons in order to illuminate the impact of DCSB on the hadron spectrum.  This analysis expands significantly on the discussion of scalar and pseudoscalar mesons in Sec.\,\ref{sec:building}.

A prediction for the spectrum follows once the gap equation's kernel is specified and the Ward-Identity solved for the Bethe-Salpeter kernel, $\Lambda$, in Fig.\,\ref{detmoldkernel}.  In the axial-vector and pseudoscalar channels, the latter is accomplished with $\Lambda_{5\beta(\mu)} = \Lambda_{5\beta(\mu)}^{\rm BC} + \Lambda_{5\beta(\mu)}^{\rm acm}$, with the explicit form of $\Lambda_{5\beta}^{\rm BC}$ given in Ref.\,\cite{Chang:2009zb}, $\Lambda_{5\beta\mu}^{\rm BC}$ constructed following the procedure described therein, and
\begin{eqnarray}
\nonumber
2 \Lambda_{5\beta(\mu)}^{\rm acm} &= & [\Gamma^{\rm acm}_{\beta}(q_{+},k_{+})+\gamma_{5}\Gamma^{\rm acm}_{\beta}(q_{-},k_{-})
\gamma_{5}] \frac{1}{S^{-1}(k_{+})+S^{-1}(-k_{-})}\Gamma_{5(\mu)}(k;P)\nonumber\\
&+&\Gamma_{5(\mu)}(q;P)\frac{1}{S^{-1}(-q_{+})+S^{-1}(q_{-})} [\gamma_{5}\Gamma^{\rm acm}_{\beta}(q_{+},k_{+})\gamma_{5}
+\Gamma^{\rm acm}_{\beta}(q_{-},k_{-})].\rule{1em}{0ex}
\end{eqnarray}
Kernels for other channels are readily constructed.

\begin{figure}[t]

\centerline{\includegraphics[clip,width=0.67\textwidth]{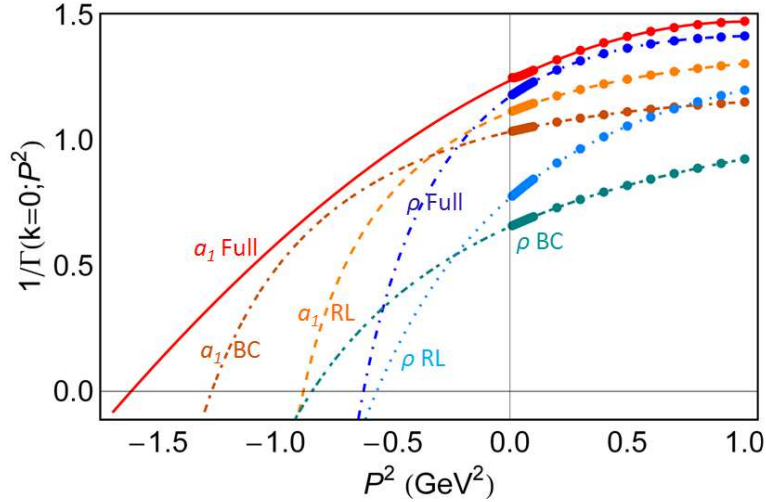}}

\caption{\label{F1}
Illustration of the procedure used to determine meson masses.
\emph{Solid curve} -- $a_1$-meson, nonperturbative kernel; \emph{dot-dash-dash} -- $a_1$, kernel derived from Eq.\,(\ref{bcvtx}) only (Ball-Chiu, BC); and \emph{dash} -- $a_1$, leading-order kernel (rainbow-ladder, RL).
\emph{Dot-dash curve} -- $\rho$-meson, nonperturbative kernel; \emph{Dot-dash-dot} -- $\rho$, BC-kernel; and \emph{dotted} -- $\rho$, RL-kernel.
\emph{Points} -- values of $1/\Gamma(k=0;P^2)$ in the given channel computed with the kernel described.  Pad\'e approximants are constructed in each case; and the location of the zero is identified with $(-m_{\rm meson}^2)$.
(Figure adapted from Ref.\,\protect\cite{Chang:2011ei}.)}
\end{figure}

Reference~\protect\cite{Chang:2011ei} computes ground-state masses using the method detailed in Ref.\,\cite{Bhagwat:2007rj}, which ensures one need only solve the gap and Bethe-Salpeter equations at spacelike momenta.  This simplifies the numerical problem.  To explain, the inhomogeneous BSE is solved for the complete Bethe-Salpeter amplitude in a particular channel on a domain of spacelike total-momenta, $P^2>0$.  Any bound-state in that channel appears as a pole in the solution at $P^2=-m_{\rm meson}^2$.  Denoting the leading Chebyshev moment of the amplitude's dominant Dirac structure by $\Gamma(k;P)$, where $k$ is the relative momentum, then $1/\Gamma(k=0;P^2)$ exhibits a zero at $(-m_{\rm meson}^2)$.  The location of that zero is determined via extrapolation of a Pad\'e approximant to the spacelike-behavior of $1/\Gamma(k=0;P^2)$.  This is illustrated for the $\rho$- and $a_1$-channels in Fig.\,\ref{F1}.

A full set of results is listed in Table~\ref{tablemasses}.
One first notes the level of agreement between Cols.~3 and 4.  This illustrates the efficacy of the extrapolation method used to compute masses: no difference is greater than 1\%.

The row associated with $m_\sigma$ is worth considering next.  First compare Cols.~1--3.  It is an algebraic result that in the RL-truncation of QCD's Dyson-Schwinger equations (DSEs), $m_\sigma \approx 2 M$, where $M$ is a constituent-like quark mass \cite{Roberts:2011cf}.  On the other hand, incorporating the quark mass function into the Bethe-Salpeter kernel via $\Gamma_\mu^{\rm BC}$ generates a strong spin-orbit interaction, which significantly boosts $m_\sigma$ (see Sec.\,\ref{sec:building}).  This feature is evidently unaffected by the inclusion of $\Gamma_\mu^{\rm acm}$; i.e., those terms associated with a dressed-quark anomalous chromomagnetic moment.
Since terms associated with pion final-state interactions are deliberately omitted in the nonperturbative kernel derived in Ref.\,\cite{Chang:2011ei}, it is noteworthy that $m_\sigma$ in Col.~1 matches estimates for the mass of the dressed-quark-core component of the $\sigma$-meson obtained using unitarised chiral perturbation theory \cite{Pelaez:2006nj,RuizdeElvira:2010cs}.  NB. In addition to providing a width, such final-state interactions necessarily reduce the real part of the mass \cite{Chang:2009ae,Cloet:2008fw}.

Now compare the entries in Rows~2, 4--6.  The $\rho$- and $a_1$-mesons have been known members of the spectrum for more than thirty years and are typically judged to be parity-partners; i.e., they would be degenerate if chiral symmetry were manifest in QCD.  Plainly, they are not, being split by more than $400\,$MeV (i.e., $> m_\rho/2$).  It is suspected that this large splitting owes to DCSB.  Hitherto, however, no symmetry-preserving treatment of bound-states could explain the splitting.  This is illustrated by Cols.~3, 4 in the Table, which show that whilst a good estimate of $m_\rho$ is readily obtained at leading-order in the systematic DSE truncation scheme of Ref.\,\cite{Bender:1996bb}, the axial-vector masses are much underestimated.  The defect persists at next-to-leading-order \cite{Watson:2004kd,Fischer:2009jm}.

\begin{table}[t]
\begin{center}
\begin{tabular*}
{\hsize}
{|l@{\extracolsep{0ptplus1fil}}
|l@{\extracolsep{0ptplus1fil}}
|l@{\extracolsep{0ptplus1fil}}
|l@{\extracolsep{0ptplus1fil}}
|l@{\extracolsep{0ptplus1fil}}|}\hline
\rule{0em}{3ex}
    & Ref.\,\protect\cite{Chang:2011ei} & Expt.~ & \emph{RL-Pad\'e}~ & \emph{RL-direct}~ \\\hline
$m_\pi$   & $0.137 $~ & 0.138 & 0.138~ & 0.137~ \\
$m_\rho$  & $0.790 \pm 0.003$~ & 0.777 & 0.754~ & 0.758~ \\
$m_\sigma$& $1.08$ & $0.4$ -- $1.2$~ & 0.645~ & 0.645~ \\
$m_{a_1}$ & $1.27 \pm 0.02$~ & $1.24 \pm 0.04$~ & 0.938~ & 0.927~  \\
$m_{b_1}$ & $1.39 \pm 0.01$~ & $1.21 \pm 0.02$~ & $0.904 $~ & 0.912~ \\
$m_{a_1}-m_\rho$~ & $0.48 \pm 0.02$ & $0.46 \pm 0.04$ & 0.18 & 0.17 \\\hline
\end{tabular*}
\end{center}
\caption{\label{tablemasses}
Computed masses.
Col.~1: Spectrum obtained with the full nonperturbative Bethe-Salpeter kernels described in Ref.\,\cite{Chang:2011ei}, which express effects of DCSB.
The method of Ref.\,\protect\cite{Bhagwat:2007rj} was used.  If noticeable, a jackknife error estimate is reported.
Col.~2 -- 
Experimental values; computed, except $m_\sigma$, from isospin mass-squared averages \protect\cite{Nakamura:2010zzi}.
%
Col.~3 -- Masses determined from the inhomogeneous BSE at leading-order in the DSE truncation scheme of Ref.\,\protect\cite{Bender:1996bb} (with this simple kernel, the jackknife error is too small to report);
and Col.~4 -- results in Ref.\,\protect\cite{Alkofer:2002bp}, obtained directly from the homogeneous BSE at the same order of truncation.
%
}
\end{table}

The analysis in Ref.\,\cite{Chang:2011ei} points to a remedy for this longstanding failure.  Using the Poincar\'e-covariant, symmetry preserving formulation of the meson bound-state problem enabled by Ref.\,\cite{Chang:2009zb}, with nonperturbative kernels for the gap and Bethe-Salpeter equations, which incorporate effects of DCSB that are impossible to capture in any step-by-step procedure for improving upon the rainbow-ladder truncation, it provides realistic estimates of axial-vector meson masses.
In obtaining these results, Ref.\,\cite{Chang:2011ei} showed that the vertex \emph{Ansatz} used most widely in studies of DCSB, $\Gamma_\mu^{BC}$, is inadequate as a tool in hadron physics.  Used alone, it increases both $m_\rho$ and $m_{a_1}$ but yields $m_{a_1}-m_\rho=0.21\,$GeV, qualitatively unchanged from the rainbow-ladder result (see Fig.\,\ref{F1}).
A good description of axial-vector mesons is only achieved by including interactions derived from $\Gamma_\mu^{\rm acm}$; i.e., connected with the dressed-quark anomalous chromomagnetic moment \cite{Chang:2010hb}.  Moreover, used alone, neither term in $\Gamma_\mu^{\rm acm}$ can produce a satisfactory result.  The full vertex \emph{Ansatz} and the associated gap and Bethe-Salpeter kernels described in Sec.\,\ref{spectrum2} are the minimum required.

Row~5 contains additional information.  The leading-covariant in the $b_1$-meson channel is $\gamma_5 k_\mu$.  The appearance of $k_\mu$ suggests immediately that dressed-quark orbital angular momentum will play a significant role in this meson's structure, even more so than in the $a_1$-channel for which the dominant covariant is $\gamma_5\gamma_\mu$.
(NB. In a simple quark-model, constituent spins are parallel within the $a_1$ but antiparallel within the $b_1$.  Constituents of the $b_1$ may therefore become closer, so that spin-orbit repulsion can exert a greater influence.)
This expectation is borne out by the following: with the full kernel, $m_{b_1}$ increases significantly more than $m_{a_1}$; and $m_{b_1}$ is far more sensitive to the interaction's momentum-space range parameter than any other state, decreasing rapidly as the interaction's spatial-variation is increasingly suppressed.
%
%

The results reviewed in this section rest on an \emph{Ansatz} for the quark-gluon vertex and whilst the best available information was used in its construction, improvement is nonetheless possible.  That will involve elucidating the role of Dirac covariants in the quark-gluon vertex which have not yet been considered and of resonant contributions; viz., meson loop effects that give widths to some of the states considered.  In cases for which empirical width-to-mass ratios are $\lesssim 25$\%, one might judge that such contributions can reliably be obtained via bound-state perturbation theory \cite{Pichowsky:1999mu}.  Contemporary studies indicate that these effects reduce bound-state masses but the reduction can uniformly be compensated by a modest inflation of the interaction's mass-scale \cite{Roberts:2011cf,Eichmann:2008ae}, so that the masses in Table~\ref{tablemasses} are semiquantitatively unchanged.  The case of the $\sigma$-meson is more complicated.  However, the prediction of a large mass for this meson's dressed-quark core leaves sufficient room for a strong reduction by resonant contributions \cite{Pelaez:2006nj,RuizdeElvira:2010cs}.

This section reviewed a continuum framework for computing and explaining the meson spectrum, which combines a veracious description of pion properties with estimates for masses of light-quark mesons heavier than $m_\rho$.
(A contemporary lattice-QCD perspective on this problem may be drawn from Ref.\,\protect\cite{Dudek:2011bn}.)
The method therefore offers the promise of a first reliable Poincar\'e-invariant, symmetry-preserving computation of the spectrum of light-quark hybrids and exotics; i.e., those putative states which are impossible to construct in a quantum mechanics based upon constituent-quark degrees-of-freedom.  So long as the promise is promptly fulfilled, the approach will provide predictions to guide the forthcoming generation of facilities.

\section{Pion Electromagnetic Form Factors}
\label{FF1}
In charting the long-range interaction between light-quarks via the feedback between experiment and theory, hadron elastic and transition form factors can provide unique information, beyond that obtained through studies of the hadron spectrum.

\subsection{Charged pion}
This is demonstrated very clearly by an analysis of the electromagnetic pion form factor, $F^{\rm em}_{\pi}(Q^2)$, because the pion has a unique place in the Standard Model.  It is a bound-state of a dressed-quark and -antiquark, and also that almost-massless collective excitation which is the Goldstone mode arising from the dynamical breaking of chiral symmetry.  This dichotomy can only be understood through the symmetry-preserving analysis of two-body bound-states \cite{Maris:1997hd}.  Furthermore, the possibility that this dichotomous nature could have wide-ranging effects on pion properties has made the empirical investigation of these properties highly desirable, despite the difficulty in preparing a system that can act as a pion target and the concomitant complexities in the interpretation of the experiments; e.g., \cite{Volmer:2000ek,Horn:2006tm,Tadevosyan:2007yd,Wijesooriya:2005ir}.

The merit of using $F^{\rm em}_{\pi}(Q^2)$ to elucidate the potential of an interplay between experiment and nonperturbative theory as a means of constraining the long-range behaviour of QCD's $\beta$-function is amplified by the existence of a prediction that $Q^2 F_{\pi}(Q^2)\approx\,$constant for $Q^2\gg m_\pi^2$ in a theory whose interaction is mediated by massless vector-bosons.  To be explicit \cite{Farrar:1979aw,Efremov:1979qk,Lepage:1980fj}:
\begin{equation}
Q^2 F_{\pi}(Q^2) \stackrel{Q^2\gg m_\pi^2}{\simeq} 16 \pi f_\pi^2 \alpha(Q^2),
\end{equation}
which takes the value $0.13\,$GeV$^2$ at $Q^2=10\,$GeV$^2$ if one uses the one-loop result $\alpha(Q^2=10\,{\rm GeV}^2)\approx 0.3$.  The verification of this prediction is a strong motivation for modern experiment \cite{Volmer:2000ek,Horn:2006tm,Tadevosyan:2007yd}, which can also be viewed as an attempt to constrain and map experimentally the pointwise behaviour of the exchange interaction that binds the pion.

Section~\ref{psmassformula} details some extraordinary consequences of DCSB, amongst them the Goldberger-Treiman relations of Eqs.\,(\ref{gtlrelE}) -- (\ref{gtlrelH}).  Of these, Eqs.\,(\ref{gtlrelF}) and (\ref{gtlrelG}) entail that the pion possesses components of pseudovector origin which alter the asymptotic form of $F_{\pi}^{\rm em}(Q^2)$ by a multiplicative factor of $Q^2$ cf.\ the result obtained in their absence \cite{Maris:1998hc}.

QCD-based DSE calculations of $F^{\rm em}_\pi(Q^2)$ exist \cite{Maris:1998hc,Maris:2000sk}, the most systematic of which \cite{Maris:2000sk} predicted the measured form factor \cite{Volmer:2000ek}.  Germane to this discourse, however, is an elucidation of the sensitivity of $F^{\rm em}_\pi(Q^2)$ to the pointwise behaviour of the interaction between quarks.  We therefore recapitulate on Refs.\,\cite{GutierrezGuerrero:2010md,Roberts:2011wy}, which explored how predictions for pion properties change if quarks interact not via massless vector-boson exchange but instead through a contact interaction; viz.,
\begin{equation}
\label{njlgluon}
g^2 D_{\mu \nu}(p-q) = \delta_{\mu \nu} \frac{1}{m_G^2}\,,
\end{equation}
where $m_G$ is a gluon mass-scale (such a scale is generated dynamically in QCD, with a value $\sim 0.5\,$GeV \cite{pepe:11,Bowman:2004jm}), and proceeded by embedding this interaction in a rainbow-ladder truncation of the DSEs.

In this case, using a confining regularisation scheme \cite{Ebert:1996vx}, the gap equation, which determines this interaction's momentum-independent dressed-quark mass, can be written
\begin{equation}
M = m +  \frac{M}{3\pi^2 m_G^2} \,{\cal C}(M^2;\tau_{\rm ir},\tau_{\rm uv})\,,
\end{equation}
where $m$ is the current-quark mass and
\begin{equation}
{\cal C}(M^2;\tau_{\rm ir},\tau_{\rm uv}) = M^2[ \Gamma(-1,M^2 \tau_{\rm uv}^2) - \Gamma(-1,M^2 \tau_{\rm ir}^2)],
\end{equation}
with $\Gamma(\alpha,y)$ being the incomplete gamma-function.  Results are presented in Table\,\ref{Table:static}.

\begin{table}[t]
\caption{Results obtained with (in GeV) $m_G=0.132\,$, $\Lambda_{\rm ir} = 0.24\,$, $\Lambda_{\rm uv}=0.905$.  Dimensioned quantities are listed in GeV.
\label{Table:static}
}
\begin{center}
\begin{tabular*}
{\hsize}
{
l@{\extracolsep{0ptplus1fil}}
|c@{\extracolsep{0ptplus1fil}}
c@{\extracolsep{0ptplus1fil}}
c@{\extracolsep{0ptplus1fil}}
|c@{\extracolsep{0ptplus1fil}}
c@{\extracolsep{0ptplus1fil}}
c@{\extracolsep{0ptplus1fil}}
c@{\extracolsep{0ptplus1fil}}
c@{\extracolsep{0ptplus1fil}}
c@{\extracolsep{0ptplus1fil}}}\hline
$m$ & $E_\pi$ & $F_\pi$ & $E_\rho$ & $M$ & $\kappa_\pi^{1/3}$ & $m_\pi$ & $m_\rho$ & $f_\pi$ & $f_\rho$ \\\hline
0 & 3.568 & 0.459 & 1.520 & 0.358 & 0.241 & 0\,~~~~~ & 0.919 & 0.100 & 0.130\rule{0ex}{2.5ex}\\
0.007 & 3.639 & 0.481 & 1.531 & 0.368 & 0.243 & 0.140 & 0.928 & 0.101 & 0.129\\\hline
\end{tabular*}
\end{center}
\end{table}

With a symmetry-preserving regularisation of the interaction in Eq.\,(\ref{njlgluon}), the Bethe-Salpeter amplitude cannot depend on relative momentum.  Hence Eq.\,(\ref{genGpi}) becomes
\begin{equation}
\Gamma_\pi(P) = \gamma_5 \left[ i E_\pi(P) + \frac{1}{M} \gamma\cdot P F_\pi(P) \right]
\end{equation}
and the explicit form of the model's ladder BSE is
\begin{equation}
\label{bsefinal0}
\left[
\begin{array}{c}
E_\pi(P)\\
F_\pi(P)
\end{array}
\right]
= \frac{1}{3\pi^2 m_G^2}
\left[
\begin{array}{cc}
{\cal K}_{EE} & {\cal K}_{EF} \\
{\cal K}_{FE} & {\cal K}_{FF}
\end{array}\right]
\left[\begin{array}{c}
E_\pi(P)\\
F_\pi(P)
\end{array}
\right],
\end{equation}
where, with $m=0=P^2$, anticipating the Goldstone character of the pion,
\begin{equation}
\begin{array}{cl}
{\cal K}_{EE}  =  {\cal C}(M^2;\tau_{\rm ir}^2,\tau_{\rm uv}^2)\,, &  {\cal K}_{EF}  =  0\,,\\
2 {\cal K}_{FE} = {\cal C}_1(M^2;\tau_{\rm ir}^2,\tau_{\rm uv}^2) \,,&
{\cal K}_{FF} = - 2 {\cal K}_{FE}\,,
\end{array}
\end{equation}
and ${\cal C}_1(z) = - z {\cal C}^\prime(z)$, where we have suppressed the dependence on $\tau_{\rm ir,uv}$.  The solution of Eq.\,(\ref{bsefinal0}) gives the pion's chiral-limit Bethe-Salpeter amplitude, which, for the computation of observables, should be normalised canonically; viz.,
\begin{equation}
%
 P_\mu = N_c\, {\rm tr} \int\! \frac{d^4q}{(2\pi)^4}\Gamma_\pi(-P)
 \frac{\partial}{\partial P_\mu} S(q+P) \, \Gamma_\pi(P)\, S(q)\,. \label{Ndef}
\end{equation}
Hence, in the chiral limit,
\begin{equation}
1 = \frac{N_c}{4\pi^2} \frac{1}{M^2} \, {\cal C}_1(M^2;\tau_{\rm ir}^2,\tau_{\rm uv}^2)
E_\pi [ E_\pi - 2 F_\pi],
\label{Norm0}
\end{equation}
and the pion's leptonic decay constant is
\begin{equation}
\label{fpi0}
f_\pi^0 = \frac{N_c}{4\pi^2} \frac{1}{M} {\cal C}_1(M^2;\tau_{\rm ir}^2,\tau_{\rm uv}^2)  [ E_\pi - 2 F_\pi ]\,.
\end{equation}

If one has preserved Eq.\,(\ref{avwtimN}), then, for $m=0$ in the neighbourhood of $P^2=0$, the solution of the axial-vector BSE has the form:
\begin{equation}
\Gamma_{5\mu}(k_+,k) = \frac{P_\mu}{P^2} \, 2 f_\pi^0 \, \Gamma_\pi(P) + \gamma_5\gamma_\mu F_R(P)
\end{equation}
and the following subset of Eqs.\,(\ref{gtlrelE}) -- (\ref{gtlrelH}) will hold:
\begin{equation}
\label{GTI}
f_\pi^0 E_\pi = M \,,\; 2\frac{F_\pi}{E_\pi}+ F_R = 1\,.
\end{equation}
Hence $F_\pi(P)$ is necessarily nonzero in a vector exchange theory, irrespective of the pointwise behaviour of the interaction.  It has a measurable impact on the value of $f_\pi$.

\begin{figure}[t] 
\centerline{\includegraphics[clip,width=0.66\textwidth]{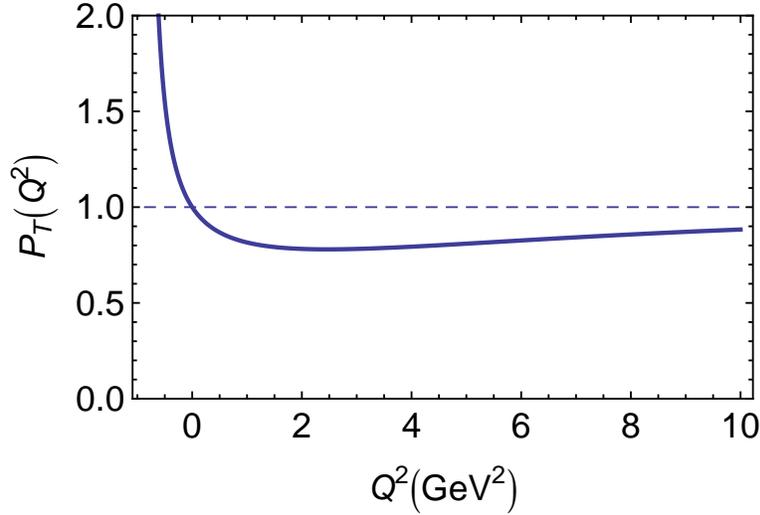}}
\caption{\label{qgammavertex}  Dressing function for the transverse piece of the quark-photon vertex, computed using the parameter values described in connection with Table~\protect\ref{Table:static}.  The pole associated with the ground-state vector meson is clear.  (Figure adapted from Ref.\,\protect\cite{Roberts:2011wy}.)}
\end{figure}

Based upon these results, one can proceed to compute the electromagnetic pion form factor in the generalised impulse approximation \cite{GutierrezGuerrero:2010md,Roberts:2011wy}; i.e., at leading-order in a symmetry-preserving DSE truncation scheme \cite{Maris:1998hc,Maris:2000sk,Roberts:1994hh}.  Namely, for an incoming pion with momentum $p_1=K-Q/2$, which absorbs a photon with space-like momentum $Q$, so that the outgoing pion has momentum $p_2=K+Q/2$,
\begin{equation}
2 K_{\mu} F_{\pi}^{\rm em}(Q^2) = 2 N_c \int\frac{d^4t}{(2\pi)^4}
{\rm tr_D} \Bigg[ i \Gamma_{\pi}(-p_2)   S(t+p_2) i \gamma_{\mu} P_{\rm T}(Q^2) S(t+p_1) \; i \Gamma_{\pi}(p_1) \; S(t) \Bigg],
\label{KF}
\end{equation}
where $P_{\rm T}(Q^2)$, depicted in Fig.\,\ref{qgammavertex}, describes the full extent of dressing on the quark-photon vertex produced by a contact interaction in the rainbow-ladder truncation \cite{Roberts:2011wy}.  The form factor is expressible as follows:
\begin{eqnarray}
F_{\pi}^{\rm em}(Q^2) &=&P_{\rm T}(Q^2)\, F_{\pi,\not \rho}^{\rm em}(Q^2),\\
F_{\pi,\not \rho}^{\rm em}(Q^2)&=& F_{\pi,EE}^{{\rm em}}(Q^2) + F_{\pi,EF}^{{\rm em}}(Q^2) + F_{\pi,FF}^{{\rm em}}(Q^2)\rule{1em}{0ex}\\
& =&  E_\pi^{\rm c}E_\pi^{\rm c}  T^{\pi}_{EE}(Q^2) + E_\pi^{\rm c} F_\pi^{\rm c} T^{\pi}_{EF}(Q^2) + F_\pi^{\rm c}F_\pi^{\rm c} T^{\pi}_{FF}(Q^2) ,
\label{F123}
\end{eqnarray}
where $F_{\pi, \not \rho}^{\rm em}(Q^2)$ is that part of the form factor produced by the undressed quark-photon vertex and the functions $T^{\pi}$ have simple algebraic forms in this calculation \cite{Roberts:2011wy}.

\begin{figure}[t] 

\begin{minipage}[t]{\textwidth}
\begin{minipage}[t]{0.4\textwidth}
\vspace*{-33ex}

\hspace*{-8em}\includegraphics[clip,height=1.1\textwidth,angle=-90]{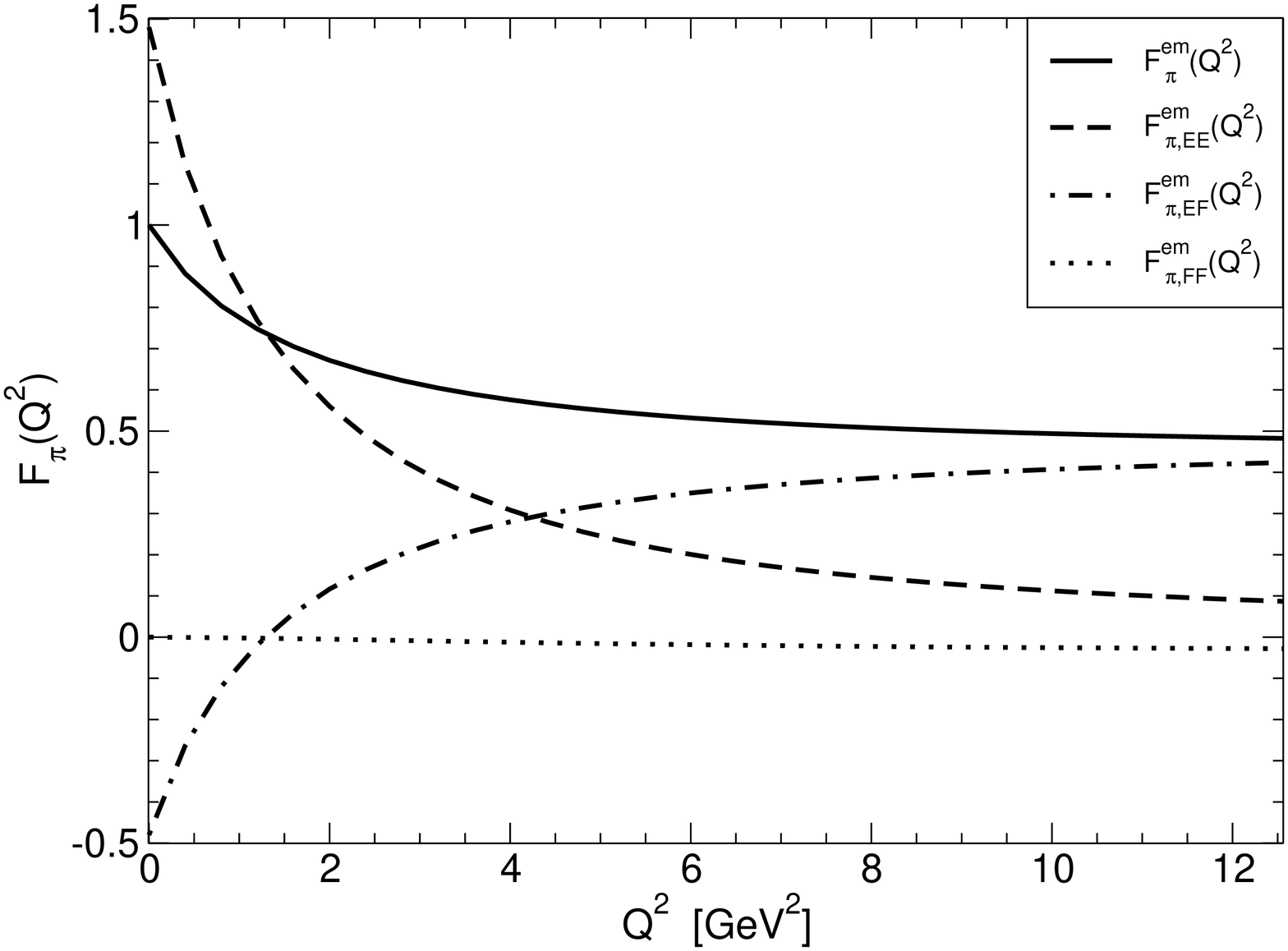}
\end{minipage}\vspace*{-23ex}
\begin{minipage}[t]{0.32\textwidth}
\hspace*{-2em}\includegraphics[clip,height=1.0\textwidth]{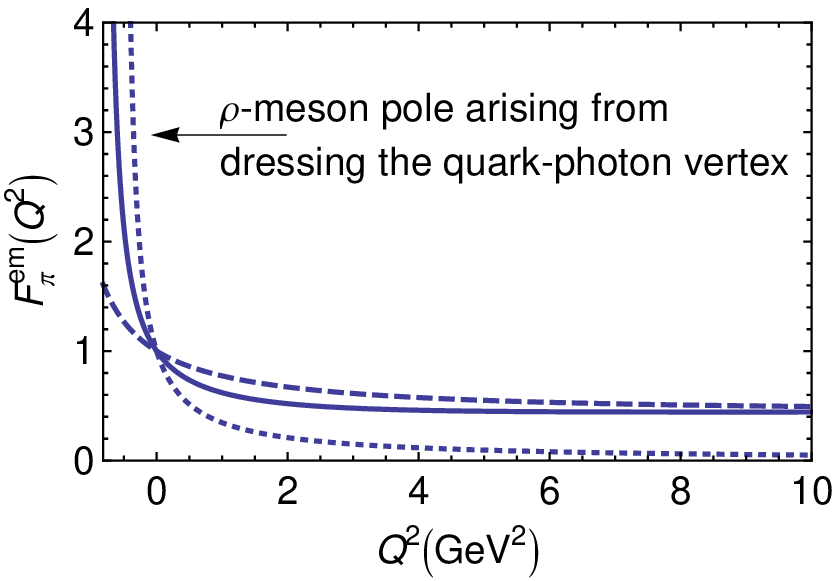}
\end{minipage}\vspace*{3ex}
\end{minipage}
\vspace*{-3ex}
%
\caption{\label{fig4}
\emph{Left panel}.
$F^{\rm em}_{\pi,\not\rho}(Q^2)$ and the separate contributions introduced in Eq.\,(\protect\ref{F123}).
$F^{\rm em}_{\pi}(Q^2=0)=1$, without fine-tuning, because a symmetry-preserving regularisation of the interaction in Eq.\,(\protect\ref{njlgluon}) was implemented.
\emph{Right panel}.
$F^{\rm em}_{\pi}(Q^2)$ computed in rainbow-ladder truncation from the interaction in Eq.\,(\protect\ref{njlgluon}): \emph{solid curve} -- fully consistent, i.e., with a dressed-quark-photon vertex so that the $\rho$-pole appears; and \emph{dashed curve} -- computed using a bare quark-photon vertex, namely, $F_{\pi,\not\rho}^{\rm em}(Q^2)$.  \emph{Dotted curve} -- fit to the result in Ref.\,\protect\cite{Maris:2000sk}, which also included a consistently-dressed quark-photon vertex and serves to illustrate the trend of contemporary data.
(Figures adapted from Refs.\,\cite{GutierrezGuerrero:2010md,Roberts:2010rn}.)
}
\end{figure}

In the left panel of Fig.\,\ref{fig4} we present $F_{\pi,\not\rho}^{\rm em}(Q^2)$ and the three separate contributions defined in Eq.\,(\ref{F123}).  It is evident that, in magnitude, $F_{\pi,EF}^{\rm em}$ contributes roughly one-third of the pion's unit charge.  This could have been anticipated from Eq.\,(\ref{Norm0}).
More dramatically, perhaps: the interaction in Eq.\,(\ref{njlgluon}) generates \begin{equation}
\label{Fpiconstant}
F_\pi^{\rm em}(Q^2 \to\infty) =\,{\rm constant.}
\end{equation}
Both results originate in the nonzero value of $F_\pi(P)$, which is a straightforward consequence of the symmetry-preserving treatment of a vector exchange theory \cite{Maris:1997hd}.  Equation~(\ref{Fpiconstant}) should not come as a surprise: with a symmetry-preserving regularisation of the interaction in Eq.\,(\ref{njlgluon}), the pion's Bethe-Salpeter amplitude cannot depend on the constituent's relative momentum.  This is characteristic of a pointlike particle, which must have a hard form factor.
The right panel of the figure illustrates that the necessary inclusion of $P_{\rm T}(Q^2)$ is critical in the timelike region and has a measurable quantitative impact for a significant range of spacelike momenta.  It does not, however, affect the truly ultraviolet behaviour.

\begin{figure}[t] 
\includegraphics[clip,height=0.50\textheight,angle=-90]{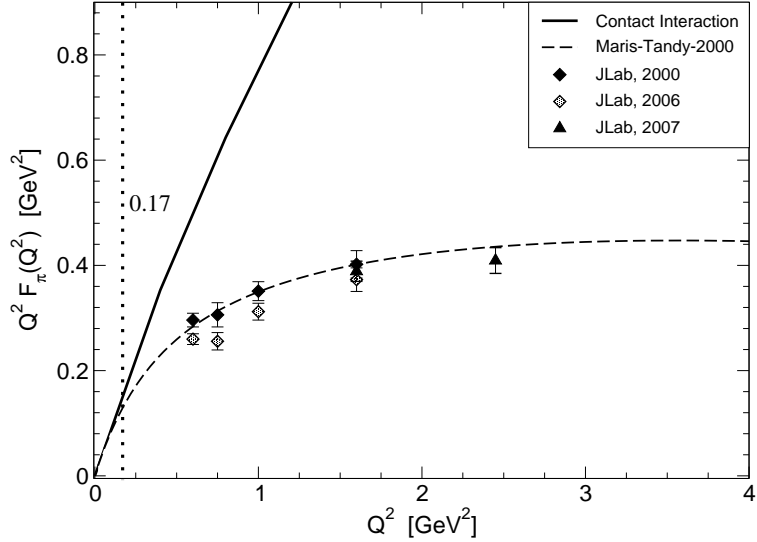}
\caption{\label{FpiUV}
\underline{Solid curve}: $Q^2 F_{\pi,\not\rho}(Q^2)$ obtained with Eq.\,(\protect\ref{njlgluon}).
%
\underline{Dashed curve}: DSE prediction \protect\cite{Maris:2000sk}, which employed a momentum-dependent renormalisation-group-improved gluon exchange interaction.
For $Q^2>0.17\,$GeV$^2\approx M^2$, marked by the vertical \emph{dotted line}, the contact interaction result for $F^{\rm em}_{\pi,\not\rho}(Q^2)$ differs from that in  Ref.\,\protect\cite{Maris:2000sk} by more than 20\%.
The data are from Refs.\,\protect\cite{Volmer:2000ek,Horn:2006tm,Tadevosyan:2007yd}.
(Figure adapted from Ref.\,\cite{GutierrezGuerrero:2010md}.)
}
\end{figure}

In Fig.\,\ref{FpiUV} we compare the form factor computed from Eq.\,(\ref{njlgluon}) with contemporary experimental data \cite{Volmer:2000ek,Horn:2006tm,Tadevosyan:2007yd} and a QCD-based DSE prediction \cite{Maris:2000sk}.  Both the QCD-based result and that obtained from the momentum-independent interaction yield the same values for the pion's static properties \cite{GutierrezGuerrero:2010md,Roberts:2010rn,Roberts:2011wy}.  However, for $Q^2>0$ the form factor computed using $\sim 1/k^2$-vector-boson exchange is immediately distinguishable empirically from that produced by a momentum-independent interaction.  Indeed, the figure shows that for $F_\pi^{\rm em}$, existing experiments can already distinguish between different possibilities for the quark-quark interaction.

Combining Figs.\,\ref{fig4} and \ref{FpiUV} it becomes apparent that $F_{\pi,EE}^{\rm em}$ is only a good approximation to the net pion form factor for $Q^2 \lsim M^2$.  $F_{\pi,EE}^{\rm em}$ and $F_{\pi,EF}^{\rm em}$ evolve with equal rapidity -- there is no reason for this to be otherwise, as they are determined by the same mass-scales -- but a nonzero constant comes quickly to dominate over a form factor that falls swiftly to zero.

It is plain now that when a momentum-independent vector-exchange interaction is regularised in a symmetry-preserving manner, the results are directly comparable with experiment, computations based on well-defined and systematically-improvable truncations of QCD's DSEs \cite{Maris:2000sk}, and perturbative QCD.  In this context it will be apparent that a contact interaction, whilst capable of describing pion static properties well, Table\,\ref{Table:static}, generates a form factor whose evolution with $Q^2$ deviates markedly from experiment for $Q^2>0.17\,$GeV$^2\approx M^2$ and produces asymptotic power-law behaviour, Eq.\,(\ref{Fpiconstant}), in serious conflict with QCD \cite{Farrar:1979aw,Efremov:1979qk,Lepage:1980fj}.

In that connection Fig.\,\ref{gluoncloud} and Eqs.\,(\ref{gtlrelE}) -- (\ref{gtlrelH}) are relevant again.
In the electromagnetic elastic scattering process, the momentum transfer, $Q$, is primarily shared equally between the pion's constituents because the bound-state Bethe-Salpeter amplitude is peaked at zero relative momentum.  Thus, one can consider $k\sim Q/2$.
The Goldberger-Trieman-like relations express a mapping between the relative momentum of the pion's constituents and the one-body momentum of dressed-quark; and the momentum dependence of the dressed-quark mass function is well-described by perturbation theory when $k^2>2\,$GeV$^2$.
Hence, one should expect a perturbative-QCD analysis of the pion form factor to be valid for $k^2=Q^2/4 \gtrsim 2\,$GeV$^2$; i.e.,
\begin{equation}
F_\pi^{\rm em}(Q^2) \approx F_\pi^{\rm em\, pQCD}(Q^2) \; \mbox{for} \; Q^2\gtrsim 8\,{\rm GeV}^2.
\end{equation}
This explains the result in Ref.\,\cite{Maris:1998hc}.
A similar argument for baryons suggests that the nucleon form factors should be perturbative for $Q^2\gtrsim 18\,$GeV$^2$.

It is worth reiterating that the contact interaction produces a momentum-independent dressed-quark mass function, in contrast to QCD-based DSE studies \cite{Roberts:2007ji,Bhagwat:2006tu} and lattice-QCD \cite{Bowman:2005vx}.  This is fundamentally the origin of the marked discrepancy between the form factor it produces and extant experiment.  Hence Refs.\,\cite{GutierrezGuerrero:2010md,Roberts:2010rn,Roberts:2011wy} highlight that form factor observables, measured at an upgraded Jefferson laboratory, e.g., are capable of mapping the running of the dressed-quark mass function.  Efforts are underway to establish the signals of the running mass in baryon elastic and transition form factors.

\subsection{Neutral pion}
The process $\gamma^\ast \gamma \to \pi^0$ is also of great interest because in order to explain the associated transition form factor within the Standard Model on the full domain of momentum transfer, one must combine, using a single internally-consistent framework, an explanation of the essentially nonperturbative Abelian anomaly with the features of perturbative QCD.  The case for attempting this received a significant boost with the publication of data from the BaBar Collaboration \cite{Aubert:2009mc} because, while they agree with earlier experiments on their common domain of squared-momentum-transfer \cite{Behrend:1990sr,Gronberg:1997fj}, the BaBar data are unexpectedly far \emph{above} the prediction of perturbative QCD at larger values of $Q^2$.

This so-called ``Babar anomaly'' was considered in Ref.\,\cite{Roberts:2010rn}, wherein it is argued that in fully-self-consistent treatments of pion: static properties; and elastic and transition form factors, the asymptotic limit of the product $Q^2 G_{\gamma^\ast\gamma \pi^0}(Q^2)$, which is determined \emph{a priori} by the interaction employed, is not exceeded at any finite value of spacelike momentum transfer: the product is a monotonically-increasing concave function.  A consistent approach is one in which: a given quark-quark scattering kernel is specified and solved in a well-defined, symmetry-preserving truncation scheme; the interaction's parameter(s) are fixed by requiring a uniformly good description of the pion's static properties; and relationships between computed quantities are faithfully maintained.

Within such an approach it is nevertheless possible for $Q^2 F^{\rm em}_\pi(Q^2)$ to exceed its perturbative-QCD asymptotic limit because the leading-order matrix-element involves two Bethe-Salpeter amplitudes.  This permits an interference between dynamically-generated infrared mass-scales in the computation.  Moreover, for $F^{\rm em}_\pi(Q^2)$ the perturbative QCD limit is more than an order-of-magnitude smaller than $m_\rho^2$.  Owing to the proximity of the $\rho$-meson pole to $Q^2=0$, the latter mass-scale must provide a fair first-estimate for the small-$Q^2$ evolution of $F^{\rm em}_\pi(Q^2)$.  A monopole based on this mass-scale exceeds the pQCD limit $\forall Q^2>0$.  For the transition form factor, however, the opposite is true because $m_\rho^2$ is less-than the pQCD limit; viz. \cite{Lepage:1980fj},
\begin{equation}
\label{BLuv}
\lim_{Q^2\to\infty} Q^2 2 G(Q^2) = 8\pi^2 f_\pi^2.
\end{equation}

\begin{figure}[t] 
\centerline{\includegraphics[clip,width=0.6\textwidth]{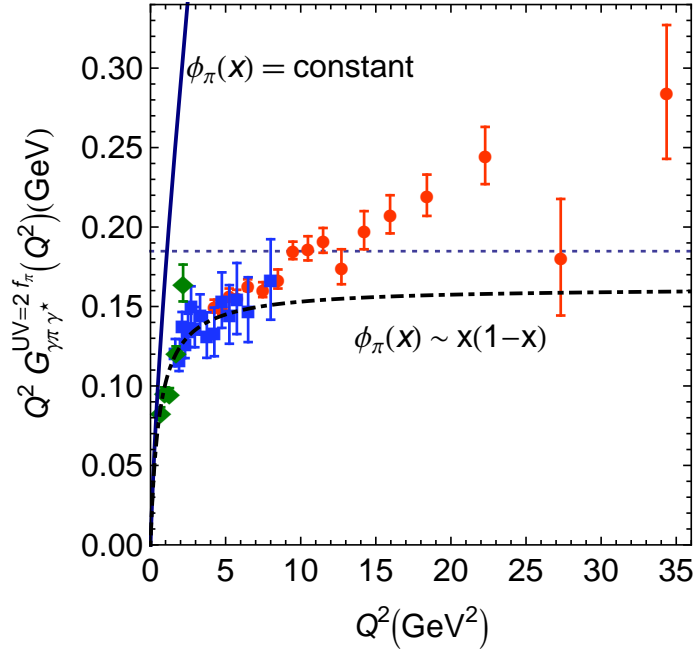}}
\caption{\label{transitionFO}
$Q^2$-weighted $\gamma^\ast \gamma \to \pi^0$ transition form factor.
Data: red circles, Ref.\,\protect\cite{Aubert:2009mc}; green diamonds, Ref.\,\protect\cite{Behrend:1990sr}; and blue squares, Ref.\protect\cite{Gronberg:1997fj}.
\emph{Solid curve} -- $Q^2 G(Q^2)$ computed using the symmetry-preserving, fully-self-consistent rainbow-ladder treatment of the contact interaction in Eq.\,(\protect\ref{njlgluon}), which produces a pion distribution amplitude $\phi_\pi(x)=\,$constant;
and \emph{dot-dashed curve} -- fit to the $\gamma^\ast\gamma \to \pi^0$ transition form factor computed in a QCD-based rainbow-ladder-truncation DSE study \cite{Maris:2002mz}.
Both curves have been divided by $(2\pi^2 f_\pi)$ in order to match the data's normalisation.
(Figure adapted from Ref.\,\protect\cite{Roberts:2010rn}.)
}
\end{figure}

The vector current-current contact-interaction canvassed in this Section may be described as a vector-boson exchange theory with vector-field propagator $(1/k^2)^\kappa$, $\kappa=0$.  It was shown \cite{Roberts:2010rn} that the consistent treatment of such an interaction produces a $\gamma^\ast\gamma \to \pi^0$ transition form factor that disagrees with \emph{all} available data.  On the other hand, precisely the same treatment of an interaction which preserves the one-loop renormalisation group behaviour of QCD, produces a form factor in good agreement with all but the large-$Q^2$ data from the BaBar Collaboration \cite{Aubert:2009mc}.  These points are illustrated in Fig.\,\ref{transitionFO}.

Studies exist which interpret the BaBar data as an indication that the pion's distribution amplitude, $\phi_\pi(x)$, deviates dramatically from its QCD asymptotic form, indeed, that $\phi_\pi(x)=\,$constant, or is at least flat and nonvanishing at $x=0,1$ \cite{Radyushkin:2009zg,Polyakov:2009je}.  However, it has often been explained \cite{Hecht:2000xa,GutierrezGuerrero:2010md,Holt:2010vj,Roberts:2010rn,Roberts:2011wy} that such a distribution amplitude characterises an essentially-pointlike pion; and, as we have seen, when used in a fully-consistent treatment, it produces results for pion elastic and transition form factors that are in striking disagreement with experiment.  Reiterating, a bound-state pion with a pointlike component will produce the hardest possible form factors; i.e., form factors which become constant at large-$Q^2$.

On the other hand, QCD-based studies produce soft pions, a valence-quark distribution amplitude for the pion that vanishes as $\sim (1-x)^2$ for $x\sim 1$, and results that agree well with the bulk of existing data.  We will return to this in the next section.

The analysis in Ref.\,\cite{Roberts:2010rn} shows that the large-$Q^2$ BaBar data is inconsistent with QCD and also inconsistent with a vector current-current contact interaction.  It supports a conclusion that the large-$Q^2$ data reported by BaBar is not a true representation of the $\gamma^\ast\gamma \to \pi^0$ transition form factor, a perspective also developed elsewhere \cite{Mikhailov:2009sa,Brodsky:2011yv,Bakulev:2011rp,Brodsky:2011xx}.  There is experimental evidence in support of this view; namely, the $\gamma^\ast \to \eta \gamma$ and $\gamma^\ast \to \eta^\prime \gamma$ transition form factors have also been measured by the BaBar Collaboration \cite{Aubert:2006cy}, at $Q^2=112\,$GeV$^2$, and in these cases the results from CLEO \cite{Gronberg:1997fj} and BaBar are fully consistent with perturbative-QCD expectations.

\section{Pion and kaon valence-quark distributions}
\label{FF2}
The past forty years have seen a tremendous effort to deduce the parton distribution
functions (PDFs) of the most accessible hadrons -- the proton, neutron and pion.  There are many reasons for this long sustained and thriving interest \cite{Holt:2010vj} but in large part it is motivated by the suspected process-independence of the usual parton distribution functions and hence an ability to unify many hadronic processes through their computation.  In connection with uncovering the essence of the strong interaction, the behaviour of the valence-quark distribution functions at large Bjorken-$x$ is most relevant.
Furthermore, an accurate determination of the behavior of distribution functions in the valence region is also important to high-energy physics.  Particle discovery experiments and Standard Model tests with colliders are only possible if the QCD background is completely understood.  QCD evolution, apparent in the so-called scaling violations by parton distribution functions, entails that with increasing center-of-mass energy, the support at large-$x$ in the distributions evolves to small-$x$ and thereby contributes materially to the collider background.
Signficantly, in the infinite momentum frame, Bjorken-$x$ measures the fraction of a hadron's four-momentum carried by the struck parton and, e.g., the valence-quark distribution function, $q_{\rm v}(x)$, measures the number-density of valence-quarks with momentum-fraction $x$.  NB.\ The nucleon PDFs are now fairly well determined for $x\lesssim 0.8$ but the pion and kaon PDFs remain poorly known on the entire domain of $x$.

Owing to the dichotomous nature of Goldstone bosons, understanding the valence-quark distribution functions in the pion and kaon is of great importance.  Moreover, given the large value of the ratio of $s$-to-$u$ current-quark masses, a comparison between the pion and kaon structure functions offers the chance to chart effects of explicit chiral symmetry breaking on the structure of would-be Goldstone modes.  There is also the prediction \cite{Ezawa:1974wm,Farrar:1975yb} that a theory in which the quarks interact via $1/k^2$-vector-boson exchange will produce valence-quark distribution functions for which
\begin{equation}
\label{pQCDuvx}
q_{\rm v}(x) \propto (1-x)^{2+\gamma} \,,\; x\gsim 0.85\,,
\end{equation}
where $\gamma\gsim 0$ is an anomalous dimension that grows with increasing momentum transfer.  (See Sec.VI.B.3 of Ref.\,\cite{Holt:2010vj} for a detailed discussion.)

Experimental knowledge of the parton structure of the pion and kaon arises primarily from pionic or kaonic Drell-Yan processes involving nucleons in heavy nuclei \cite{Wijesooriya:2005ir,Badier:1980jq,Badier:1983mj,Betev:1985pg,Conway:1989fs}.  Theoretically, given that DCSB plays a crucial role in connection with pseudoscalar mesons, one must employ an approach that realistically expresses this phenomenon.  The DSEs therefore provide a natural framework: studies of the pion and kaon exist and will be reviewed here.  The first \cite{Hecht:2000xa} computed pion PDFs, using efficacious parametrisations of both the Bethe-Salpeter amplitude and dressed-quark propagators \cite{Roberts:1994hh,Burden:1995ve,ElBennich:2010ha}.  The second \cite{Nguyen:2011jy} employed direct, numerical DSE solutions in the computation of the pion and kaon PDFs, adapting the approach employed in successful predictions of electromagnetic form factors \cite{Maris:1999bh,Maris:2000sk,Maris:2002mz,Holl:2005vu}; and also studied the ratio $u_K(x)/u_\pi(x)$ in order to elucidate aspects of the influence of an hadronic environment.

In rainbow-ladder truncation, one obtains the pion's valence-quark distribution from
\begin{eqnarray}
u_\pi(x) = -\frac{1}{2} \int \frac{d^4 \ell}{(2\pi)^4}  {\rm tr}_{\rm cd}\, \left[ \Gamma_\pi(\ell,-P) \,S_u(\ell)\, \Gamma^n(\ell;x) \, S_u(\ell)\, \Gamma_\pi(\ell,P)\, S_d(\ell-P) \right] ,
\label{Eucl_pdf_LR_Ward}
\end{eqnarray}
wherein the Bethe-Salpeter amplitudes and dressed-quark propagators are discussed above and $\Gamma^n(\ell;x)$ is a generalization of the dressed-quark-photon vertex, describing a dressed-quark scattering from a zero momentum photon.  It satisfies a BSE (here with a rainbow-ladder kernel) with the inhomogeneous term  $i\gamma\cdot n \, \delta(\ell \cdot n - x P\cdot n)$.  Here $n_\mu$ is a  light-like vector satisfying \mbox{$n^2 = 0$}.
In choosing rainbow-ladder truncation one implements a precise parallel to the symmetry-preserving treatment of the pion charge form factor at \mbox{$Q^2 = 0 $}, wherein the vector current is conserved by use of ladder dynamics at all three vertices and rainbow dynamics for all three quark propagators \cite{Roberts:1994hh,Maris:1998hc,Maris:1999bh,Maris:2000sk}.   Equation~(\ref{Eucl_pdf_LR_Ward}) ensures automatically that
\begin{equation}
 \langle x_f^0 \rangle := \int_0^1 dx \,  q^v_f(x) = 1\; \mbox{for}\; f = u, \bar d\,,
\end{equation}
since $ \int dx \, \Gamma^n(\ell;x)$ gives the Ward-identity vertex and the Bethe-Salpeter amplitudes are canonically normalised.

\begin{figure}[t]
\includegraphics[clip,width=0.66\textwidth]{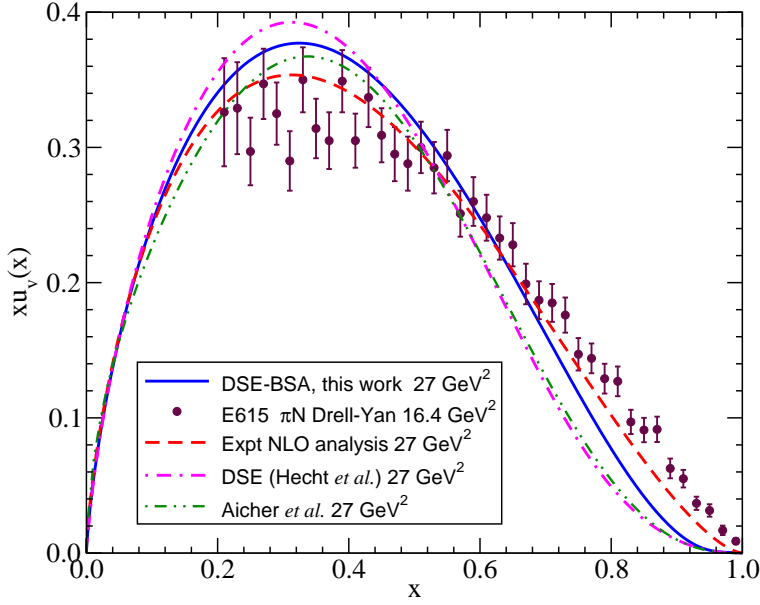}

\caption{ Pion valence quark distribution function evolved to (5.2~GeV)$^2$.  \emph{Solid curve} -- full DSE calculation \protect\cite{Nguyen:2011jy}; \emph{dot-dashed curve} -- semi-phenomenological DSE-based calculation in Ref.\,\protect\cite{Hecht:2000xa}; \emph{filled circles} -- experimental data from Ref.\,\protect\cite{Conway:1989fs}, at scale (4.05\,{\rm GeV})$^2$;
\emph{dashed curve} -- NLO reanalysis of the experimental data \protect\cite{Wijesooriya:2005ir};
and \emph{dot-dot-dashed curve} -- NLO reanalysis of experimental data with inclusion of soft-gluon resummation \protect\cite{Aicher:2010cb}.
(Adapted from Ref.\,\protect\cite{Nguyen:2011jy}.) \label{fig:pi_DSE}}
\end{figure}

Figure~\ref{fig:pi_DSE} displays the DSE results for the pion's valence $u$-quark distribution, evolved from a resolving scale $Q_0^2=(0.6\,$GeV)$^2$ to $Q^2 = (5.2~{\rm GeV})^2$ using leading-order evolution, and a comparison with $\pi N$ Drell-Yan data \cite{Conway:1989fs} at a scale  $Q^2 \sim (4.05~{\rm GeV})^2$, inferred via a leading-order analysis.  The computation's resolving scale, $Q_0$, was fixed by matching the $\langle x^n\rangle^\pi$ moments for $n=1,2,3$ to an experimental analysis at (2\,{\rm GeV})$^2$ \cite{Sutton:1991ay}.

It is notable that at $Q_0$ the DSE results yield
\begin{equation}
\label{momcons}
2 \, \langle x \rangle^\pi_{Q_0} = 0.7\,,\;
2 \, \langle x \rangle^K_{Q_0} = 0.8\,.
\end{equation}
(For comparison, the parametrised valence-like pion parton distributions of Ref.\,\protect\cite{Gluck:1998xa} yield a gluon momentum fraction of $\langle x_g\rangle^\pi_{Q_0=0.51} = 0.3$.)
In each case the remainder of the hadron's momentum is carried by gluons, which effect binding of the meson bound state and are invisible to the electromagnetic probe.  Some fraction of the hadron's momentum is carried by gluons at all resolving scales unless the hadron is a point particle \cite{Holt:2010vj}.  Indeed, it is a simple algebraic exercise to demonstrate that the only non-increasing, convex function which can produce $\langle x^0\rangle =1$ and $\langle x \rangle = \frac{1}{2}$, is the distribution $u(x)=1$, which is uniquely connected with a pointlike meson; viz., a meson whose Bethe-Salpeter amplitude is momentum-independent.  Thus Eqs.\,(\ref{momcons}) are an essential consequence of momentum conservation.

Whilst the DSE results in Fig.\,\ref{fig:pi_DSE} are both consistent with Eq.\,(\ref{pQCDuvx}); i.e., they produce algebraically the precise behaviour predicted by perturbative QCD, on the valence-quark domain it is evident that they disagree markedly with the Drell-Yan data reported in Ref.\,\cite{Conway:1989fs}.  This tension was long seen as a crucial mystery for a QCD description of the lightest and subtlest hadron \cite{Holt:2010vj}.  Its re-emergence with Ref.\,\cite{Hecht:2000xa} motivated a NLO reanalysis of the Drell-Yan data \cite{Wijesooriya:2005ir}, the result of which is also displayed in Fig.\,\ref{fig:pi_DSE}.  At NLO the extracted PDF is softer at high-$x$ but the discrepancy nevertheless remains.
To be precise, Ref.\,\cite{Wijesooriya:2005ir} determined a high-$x$ exponent of $\beta \simeq 1.5$ whereas the exponents produced by the DSE studies \cite{Nguyen:2011jy,Hecht:2000xa} are, respectively, $2.1$ and $2.4$ at the common model scale.  They do not allow much room for a harder PDF at high-$x$.

Following the highlighting of this discrepancy in Ref.\,\cite{Holt:2010vj}, a resolution of the conflict between data and well-constrained theory was proposed.  In Ref.\,\cite{Aicher:2010cb} a long-overlooked effect was incorporated; namely, ``soft-gluon resummation.'' With the inclusion of this next-to-leading-logarithmic threshold resummation effect in the calculation of the Drell-Yan cross section, a considerably softer valence-quark distribution was obtained at high-$x$.
This is readily understood.  The Drell-Yan cross-section factorises into two pieces: one hard and the other soft.  The soft piece involves the PDF and the hard piece is calculable in perturbation theory.  Adding additional interactions to the latter, which are important at large-$x$; viz., soft gluons, provides greater strength in the hard piece on the valence-quark domain.  Hence a description of the data is obtained with a softer PDF.
Indeed, the distribution obtained thereby matches precisely the expectations based on perturbative-QCD and obtained using DSEs.  This is evident in a comparison between the \emph{dash-dot} and \emph{dash-dot-dot} curves in Fig.\,\ref{fig:pi_DSE}.

\begin{figure}[t]
\includegraphics[clip,width=0.66\textwidth]{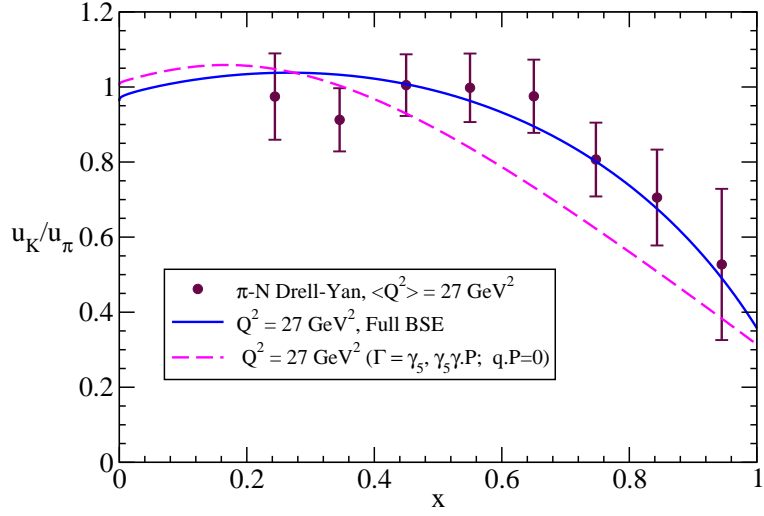}

\caption{\label{fig:pi_DSE_ratio}
DSE prediction for the ratio of $u$-quark distributions in the kaon and pion \protect\cite{Holt:2010vj,Nguyen:2011jy}.  The full Bethe-Salpeter amplitude produces the \emph{solid} curve; a reduced BSE vertex produces the \emph{dashed} curve.  The reduced amplitude retains only the invariants and amplitudes  involving pseudoscalar and axial vector Dirac matrices, and ignores dependence on the variable $q\cdot P$.  These are part of the reductions that occur in a pointlike treatment of pseudoscalar mesons.  The experimental data is from \protect\cite{Badier:1980jq,Badier:1983mj}.
(Adapted from Ref.\,\protect\cite{Nguyen:2011jy}.)}
\end{figure}

The ratio $u_K/u_\pi$ measures the effect of the local hadronic environment.  In the kaon, the $u$-quark is bound with a heavier partner than in the pion and this should cause $u_K(x)$ to peak at lower-$x$ than $u_\pi(x)$.  The DSE prediction \cite{Holt:2010vj,Nguyen:2011jy} is shown in Fig.\,\ref{fig:pi_DSE_ratio} along with available Drell-Yan data \cite{Badier:1980jq,Badier:1983mj}.  The parameter-free DSE result agrees well with the data.
We note that
\begin{equation}
\frac{u_K(0)}{u_\pi(0)} \stackrel{\mbox{\footnotesize\rm DGLAP:}Q^2\to \infty}{\to} 1\,;
\end{equation}
viz, the ratio approaches one under evolution to larger resolving scales owing to the increasingly large population of sea-quarks produced thereby \cite{Chang:2010xs}.  On the other hand, the value at $x=1$ is a fixed-point under evolution:
\begin{equation}
\forall Q_1^2>Q_0^2,\;
\left.\frac{u_K(1)}{u_\pi(1)}\right|_{Q_1^2} \stackrel{\mbox{\footnotesize\rm DGLAP:}Q_0^2\to  Q_1^2}{=}  
\left. \frac{u_K(1)}{u_\pi(1)} \right|_{Q_0^2}=
\left. \frac{u_K(1)}{u_\pi(1)} \right|_{Q_0^2}
\end{equation}
i.e., it is the same at every value of the resolving scale $Q^2$, and is therefore a persistent probe of nonperturbative dynamics \cite{Holt:2010vj}.

With Ref.\,\cite{Nguyen:2011jy} a significant milestone was achieved; viz., unification of the computation of distribution functions that arise in analyses of deep inelastic scattering with that of numerous other properties of pseudoscalar mesons, including meson-meson scattering \cite{Bicudo:2001aw,Bicudo:2001jq} and the successful prediction of electromagnetic elastic and transition form factors.  The results confirm the large-$x$ behavior of distribution functions predicted by the QCD parton model; provide a good account of the $\pi$-$N$ Drell-Yan data for $u_\pi(x)$; and a parameter-free prediction for the ratio $u_K(x)/u_\pi(x)$ that agrees with extant data, showing a strong environment-dependence of the $u$-quark distribution.  The new Drell-Yan experiment running at FNAL is capable of validating this comparison, as is the COMPASS~II experiment at CERN.  Such an experiment should be done so that complete understanding of QCD's Goldstone modes can be claimed.

\section{Baryon Properties}
\label{FF3}
While a symmetry-preserving description of mesons is essential, it is only part of the physics that nonperturbative QCD must describe because Nature also presents us with baryons: light-quarks in three-particle composites.  An explanation of the spectrum of baryons and the nature of interactions between them is basic to understanding the Standard Model.  The present and planned experimental programmes at JLab, and other facilities worldwide, are critical elements in this effort.

No approach to QCD is comprehensive if it cannot provide a unified explanation of both mesons and baryons.  We have explained that DCSB is a keystone of the Standard Model, which is evident in the momentum-dependence of the dressed-quark mass function -- Fig.\,\ref{gluoncloud}: it is just as important to baryons as it is to mesons.  The DSEs furnish the only extant framework that can simultaneously connect both meson and baryon observables with this basic feature of QCD, having provided, e.g., a direct correlation of meson and baryon properties via a single interaction kernel, which preserves QCD's one-loop renormalisation group behaviour and can systematically be improved \cite{Eichmann:2008ae,Eichmann:2008ef,Eichmann:2011vu,Mader:2011zf}.

In quantum field theory a baryon appears as a pole in a six-point quark Green function.  The residue is proportional to the baryon's Faddeev amplitude, which is obtained from a Poincar\'e covariant Faddeev equation that sums all possible exchanges and interactions that can take place between three dressed-quarks.  A tractable Faddeev equation for baryons \cite{Cahill:1988dx} is founded on the observation that an interaction which describes colour-singlet mesons also generates nonpointlike quark-quark (diquark) correlations in the colour-$\bar 3$ (antitriplet) channel \cite{Cahill:1987qr}.  The dominant correlations for ground state octet and decuplet baryons are scalar ($0^+$) and axial-vector ($1^+$) diquarks because, for example, the associated mass-scales are smaller than the baryons' masses \cite{Burden:1996nh,Maris:2002yu} and their parity matches that of these baryons.  It follows that only they need be retained in approximating the quark-quark scattering matrix which appears as part of the Faddeev equation \cite{Cloet:2008re,Roberts:2011cf,Eichmann:2008ef}.

We note that diquarks do not appear in the strong interaction spectrum \cite{Bender:1996bb,detmold,Bhagwat:2004hn} but the attraction between quarks in this channel justifies a picture of baryons in which two quarks within a baryon are always correlated as a colour-$\bar 3$ diquark pseudoparticle, and binding is effected by the iterated exchange of roles between the bystander and diquark-participant quarks.   Here it is important to emphasise strongly that QCD supports \emph{nonpointlike} diquark correlations \cite{Roberts:2011wy,Maris:2004bp}.  Hence models that employ pointlike diquark degrees of freedom have little connection with QCD.

Numerous properties of the nucleon have recently been described using the Faddeev equation just outlined,
e.g.: a survey of the form factors \cite{Cloet:2008re}, with additional results explicated in Refs.\,\cite{Chang:2009ae,Cloet:2008wg,Cloet:2008fw,Chang:2010jq,Roberts:2010hu,
Chang:2011tx,Cloet:2011qu};
indications of the role of quark orbital angular momentum in forming the nucleon's spin \cite{Cloet:2007pi}; and computation of the nucleon's $\sigma$-term \cite{Flambaum:2005kc}.
This body of work will be reviewed elsewhere.
However, much more can and should be done, as described, for example, in Sec.\,III of Ref.\,\cite{Aznauryan:2009da}; and Ref.\,\cite{Roberts:2011cf} is a first step.

It is expected that feedback between DSE results and extant and forthcoming precision data on nucleon elastic and transition form factors can serve as a practical means by which to empirically chart the momentum evolution of the dressed-quark mass function, and therefrom the infrared behavior of QCD's $\beta$-function.  In particular, it should enable the unambiguous location of the transition boundary between the constituent- and current-quark domains, signalled by the sharp drop with increasing momentum that is apparent in Fig.\,\ref{gluoncloud}, which can likely be related to an infrared inflexion point in QCD's running coupling, whose properties are determined by the $\beta$-function.

\section{Epilogue}
Dynamical chiral symmetry breaking (DCSB) is a fact in QCD.  It is manifest in dressed-propagators and vertices, and, amongst other things, it is responsible for:
the transformation of the light current-quarks in QCD's Lagrangian into heavy constituent-like quark's, in terms of which order was first brought to the hadron spectrum;
the unnaturally small values of the masses of light-quark pseudoscalar mesons;
the unnaturally strong coupling of pseudoscalar mesons to light-quarks -- $g_{\pi \bar q q} \approx 4.3$;
and the unnaturally strong coupling of pseudoscalar mesons to the lightest baryons -- $g_{\pi \bar N N} \approx 12.8 \approx 3 g_{\pi \bar q q}$.

Herein we have used a diverse range of phenomena to illustrate the dramatic impact that DCSB has upon meson observables, amongst them: the spectrum; form factors; and parton distribution functions.  Unavoidably, many valuable contributions during the last few years were overlooked, some of which may be pursued through Refs.\,\cite{Pennington:2010gy,Blank:2011qk}.
We have also completely neglected an extensive body of work that focuses on elucidating the nature of the primordial state of matter that has been recreated with RHIC:
Refs.\,\cite{Qin:2010nq,Qin:2010pc,Liu:2011zz,Blank:2011qk,He:2008yr} serve as a starting point for exploring this area of DSE research.
(As noted in Ref.\,\protect\cite{Maris:2000ig}, the rainbow truncation of QCD's gap equation excludes what in a point-meson field theory would appear as $1/N_c$-suppressed meson-loop dressing of the quark propagator, and this is why chiral symmetry restoration is a mean field transition in all models within the rainbow-truncation class \protect\cite{Holl:1998qs}.  Such meson-loop corrections are capable of correcting the critical exponents; i.e., of introducing that infrared divergence in the dressed-quark self-energy which can force a deviation from mean field critical exponents in the chiral symmetry restoring transition.  This is illustrated in Ref.\,\protect\cite{Fischer:2011pk}.)
In addition, there are attempts to derive an equation of state that may be used in modern astrophysics in order, e.g., to assist in the identification of a neutron star with a quark-matter core \cite{Klahn:2009mb,Li:2011vd}.

A ``smoking gun'' for DCSB is the behaviour of the dressed-quark mass function.  The momentum dependence manifest in Fig.\,\ref{gluoncloud} is an essentially quantum field theoretical effect.  Exposing and elucidating its consequences therefore requires a nonperturbative and symmetry-preserving approach, where the latter means preserving Poincar\'e covariance, chiral and electromagnetic current-conservation, etc.  The Dyson-Schwinger equations (DSEs) provide such a framework.  Experimental and theoretical studies are underway that will identify observable signals of $M(p^2)$ and thereby confirm and explain the mechanism responsible for the vast bulk of visible mass in the Universe.

There are many reasons why this is an exciting time in hadron physics.  We have focused on one.  Namely, through the DSEs, we are positioned to unify phenomena as apparently diverse as: the hadron spectrum; hadron elastic and transition form factors, from small- to large-$Q^2$; and parton distribution functions.  The key is an understanding of both the fundamental origin of nuclear mass and the far-reaching consequences of the mechanism responsible; namely, DCSB.  These things might lead us to an explanation of confinement, the phenomenon that makes nonperturbative QCD the most interesting piece of the Standard Model.


\section*{\large Acknowledgements}
We acknowledge valuable input from A.~Bashir, I.\,C.~Clo\"et, B.~El-Bennich, L.~X.~Guti\'errez-Guerrero, A.~K{\i}z{\i}lers\"u, Y.-X.~Liu, S.-x.~Qin, J.~Rodriguez-Quintero, S.\,M.~Schmidt and D.~J.~Wilson.
This work was supported by:
the U.\,S.\ Department of Energy, Office of Nuclear Physics, contract no.~DE-AC02-06CH11357;
Forschungszentrum J\"ulich GmbH;
and the U.\,S.\ National Science Foundation, under grant no.\ PHY-0903991, in conjunction with a CONACyT Mexico-USA collaboration grant.



\bibliographystyle{aipproc}   

\end{document}